\documentclass[pdflatex,sn-mathphys-num]{sn-jnl}

\usepackage{amsmath,amsfonts,amssymb}
\usepackage{mathrsfs,bbm}
\usepackage{hyperref} 
\usepackage{cleveref}
\usepackage[normalem]{ulem} 

\def\p{\partial}

\def\non{\nonumber}

\begin{document}

\title{A 3D Summation-by-Parts scheme on a Hyperboloidal Foliation of Minkowski}

\author[1,2]{\fnm{Shalabh} \sur{Gautam}}\email{shalabhgautam@bimsa.cn}

\affil[1]{
 \orgname{Beijing Institute of Mathematical Sciences and Applications (BIMSA)}, \orgaddress{\street{No. 544, Hefangkou Village Huaibei Town}, \city{Huairou District}, \postcode{101408}, \state{Beijing}, \country{China}}}

\affil[2]{
 \orgname{Yau Mathematical Sciences Center~(YMSC)}, \orgaddress{\street{Jingzhai, Tsinghua University}, \city{Haidian District}, \postcode{100084}, \state{Beijing}, \country{China}}}

\date{\today}

\abstract{
This paper summarises our previous work on a fully~$3$D Summation-by-Parts scheme, derived for a class of linear wave equations on hyperboloidal slices on a fixed Minkowski background. The scheme is derived in spherical polar coordinates, and allows having grid points at the origin and on the~$z$-axis, despite coordinate singularities, and at infinity, by introducing compactification followed by rescaling, and is proved to be stable. Reducing it to the standard Cauchy problem, or to finite spacelike slices with an outer boundary, will follow a similarly. Second-order accurate finite-difference methods are used to implement this scheme numerically, but higher-order finite-difference or spectral methods could also be used. Kreiss-Oliger dissipation operators~(KODOs) are generalized to curvilinear coordinates and are defined everywhere in the domain, including at the boundary points, such that they satisfy the dissipative property~(DP) in the energy norms. We also propose new norm convergence tests that include all the grid points at all resolutions and produce more accurate results. Promising results are obtained, giving hope for application to fully nonlinear systems, like the Einstein Field Equations, and extracting the resulting gravitational waves free of systematic errors or gauge ambiguities.
}

\keywords{Summation-by-parts, hyperboloidal slices, general relativity, gravitational waves, numerical relativity}

\maketitle

\section{Introduction}\label{sec:intro}

The ultimate goal of numerical relativity~(NR) is to solve the Einstein field equations~(EFEs) numerically for any initial data~(ID) of interest at any desired accuracy and up to any desired time. Including the entire spacetime backreaction effects and producing highly accurate gravitational waves~(GW) requires including future null infinity, denoted by~$\mathscr{I}^+$, in the computational domain with a well-posed formulation of the~EFEs. As suggested by Penrose~\cite{Pen63},~$\mathscr{I}^+$ is the set of endpoints of all future-directed null geodesics in the exterior of a given spacetime. This problem needs to be addressed geometrically, analytically, and numerically.

The geometric part of the problem involves introducing a spacetime foliation that maps~$\mathscr{I}^+$ onto the computational grid. The state-of-the-art methods are Cauchy-Characteristic Extraction~(CCE)~\cite{BisGomLeh96, BisGomLeh97, MoxSchTeu21}, and Cauchy-Characteristic Matching~(CCM)~\cite{Win12, MaMoxSch23, MaSchMox24}, and an approximate method of extrapolation~\cite{BoyMro09}. However,~CCM has issues with well-posedness~\cite{GiaHilZil20, GiaBisHil22, GiaBisHil23},~CCE does not feedback backreaction effects from the characteristic region to the Cauchy domain. In both cases, the solutions remain gauge dependent. Additionaly, both~CCE and extrapolation methods suffer from outer boundary effects, and the latter introduces systematic errors due to extrapolation in the resulting signals.

To address these limitations, Friedrich~\cite{Fri81, Fri81a, Fri83} proposed a hyperboloidal foliation approach, together with the conformal~EFEs~(cEFEs), which use conformal compactification suggested by Penrose~\cite{Pen63} in~$1963$. The initial value problems associated with this formulation were further investigated by Andersson et al~\cite{AndChrFri92, AndChr93, AndChr94}, while the first numerical implementations were carried out by H\"ubner~\cite{Hub94, Hub99, Hub99a, Hub01} and Frauendiener~\cite{Fra97, Fra97a, Fra98}. Despite being completely regular, the potential drawback of this formulation is that~$\mathscr{I}^+$ here moves along the incoming null direction and collapses to a single grid point, corresponding to the future timelike infinity~$i^+$, in a finite conformal time, imposing limitations on the overall numerical accuracy of the system.

One solution to this problem is introducing~$\mathscr{I}^+$-fixing proposed by Zengino{\u{g}}lu~\cite{Zen07, Zen08} along with an uncompactified time coordinate, so one can attain long-time evolutions without losing computational resolution~\cite{ZenKid10, Zen11, ZenGal12, YanZimZen13, Rin25}. However, none of the formulations with~$\mathscr{I}^+$-fixing gives a completely regular system of equations~\cite{MonRin08, Rin09, Rin10, RinMon13, Rin14, BaaRin16, MalRin18, VanHusHil14, VanHus14, Van15, VanHus16, VanHus17, Van23, Van23a, VanVal24, AlvVan25, AlvVan25a, Hil15, HilHarBug16, GasHil19, GasGauHil20, DuaFenGas21, DuaFenGas22, DuaFenGas22a, PetGauRai23, PetGauVan24, NakNakRac17, CsuRac19, CsuRac23, CsuRac25}. Nonetheless, it is hoped that this analytical aspect of the problem will be resolved in the near future.
 
This paper summarises our previous works~\cite{GauVanHil21, GauRedKum26} that address the numerical part of the problem by deriving a suitable discretization scheme for a regularized system on hyperboloidal slices. While this aspect requires exploration even for the simplest cases, such as the linear wave equation~(LWE), it establishes a necessary framework for generalizing to a fully nonlinear system of the~EFEs in generalized harmonic gauge~(GHG)~\cite{Fri85, Gar02} in full~$3$D. Our summation-by-parts~(SBP) scheme simplifies the numerical implementation of a hyperbolic system by:
\begin{enumerate}
\item {\it Handling Coordinate Singularities:} It allows having grid points directly at the origin and on the~$z$-axis without requiring intricate coordinate transformations or manual regularization of the evolution variables at these points.
\item {\it Stable Boundary Inclusion:} The approach offers an energy-stable treatment of~$\mathscr{I}^+$, enabling the unambiguous extraction of~GW within the computational domain, eliminating systematic errors of extrapolation and gauge ambiguities.
\item {\it Constraint Damping Integration:} It introduces a geometric way for adding first-order reduction~(FOR) constraint damping terms, which generalizes directly to the quasilinear hyperbolic systems essential for long-term stability and consistency.
\item {\it Artificial Dissipation:} It proposes a covariant approach for deriving artificial dissipation operators and a geometric way of introducing them to the associated discrete system, effectively eliminating the numerical noise in our coordinate basis.
\end{enumerate}

The~SBP scheme~\cite{Str94} has proven to be a very powerful tool in various fields of computational physics and mathematics~\cite{VreSch26, HalHarNch26, ManMalNch26, SteDur26, WorFerZin26, GlaIskLam26, RicLeeDur24, SteLeeDur24, GlaRanHen25, RanWinSch23, SarTig12}. When expressed in spherical polar coordinates within a Minkowski spacetime, the~LWE encounters coordinate singularities at the origin and along the~$z$-axis. In general~$d+2$ dimensions, consisting of a~$d$-dimensional sphere along with time and radial directions, the standard~LWE presents polar singularities along~$d$ axes. This issue was addressed in~\cite{GunGarGar10, CsiLasRac12} by reducing it to an effective spherically symmetric problem using~$d$-dimensional spherical harmonics, and then overcoming the singularity at the origin. These spherical modes were then superposed analytically to derive the general solution. The resulting spherically symmetric system was subsequently generalized to hyperboloidal slices in~\cite{GauVanHil21}.

This paper presents a full~$3$D spatial discretization without introducing any such decomposition, which can be generalized to a~$d+2$ dimensional Minkowski spacetime in a similar manner. Although we use the standard~RK4 time integrator, assuming that it satisfies certain required properties of continuum time integration, more rigorous integrators can also be employed as given in~\cite{Tad23, MarBoyBru14, MarBoyGle19, BoyMarSil22, MarBraZen23, BoyMar23, CorCerIba12}. Likewise, we implement second-order accurate finite-difference~(FD) methods, but higher-order accurate~FD operators or pseudo-spectral methods could also be used.

We use uppercase Latin letters with primed indices, represented as~$X^{\mu'} = (T,R,\theta,\phi)$, to denote the standard Cauchy coordinates, and lowercase with unprimed indices,~$x^\mu = (t,r,\theta,\phi)$, for the hyperboloidal ones. Throughout the presentation, lowercase Greek indices,~$\mu, \nu, \ldots$, will represent spacetime indices~$(0,1,2,3)$, the latter half of the Latin indices,~$i, j, k, \ldots$ will denote spatial components~$(1,2,3)$, while the first half of lowercase Latin indices~$a, b, c, \ldots$ will denote the abstract indices. We will adopt the signature~$(-,+,+,+)$ for the spacetime metric.

This presentation is structured as follows. Section~\ref{sec:continuum_setup} derives the continuum set of equations, which are discretized in Sec.~\ref{sec:Discretization} and then implemented numerically for some specific cases in Sec.~\ref{sec:numerics}. We conclude with results and outlook in Sec.~\ref{sec:conclusions}.

\section{Continuum Setup}\label{sec:continuum_setup}

This section introduces a hyperboloidal foliation of Minkowski spacetime, casts the evolution equations on these slices, and regularizes them at coordinate singularities as well as at~$\mathscr{I}^+$, and then presents the associated conserved energy. We begin with the Minkowski metric represented in standard spherical polar coordinates~$X^{\mu'} = (T,R,\theta,\phi)$
\begin{align}\label{eq:MK_metric}
& ds^2 = -dT^2 + dR^2 + R^2 \, d\theta^2 + R^2 \, \sin^2 \theta \, d\phi^2 \, , \non \\
\textrm{with} \quad & T \in (-\infty,\infty) \, , \; R \in [0,\infty) \, , \; \theta \in [0,\pi] \, , \; \textrm{and } \phi \in [0, 2\pi) \, ,
\end{align}
which gives the outgoing and incoming characteristic speeds to be~$C^R_\pm = \pm 1$. As the null signals reach~$\mathscr{I}^+$ in their infinite future, whereas the Cauchy slices reach spacelike infinity,~$i^0$, as~$R \rightarrow \infty$, a foliation defined by the level sets of~$T$ cannot extract signal at~$\mathscr{I}^+$. We therefore introduce the hyperboloidal foliation described below.

\subsection{Hyperboloidal Slices}\label{sec:Hyperboloidal_slices}

The hyperboloidal slices are the ones that are spacelike everywhere but reach~$\mathscr{I}^+$ as~$R \rightarrow \infty$. We obtain them here by introducing the height function~$H(R)$~\cite{BeiMur98, MalMur03, Zen07}
\begin{equation}\label{eq:Hyperboloidal_Coords}
T = t + H(R) \, , \quad \text{with} \, , \quad R = R(r) \, .
\end{equation}
Here~$t$ is called the hyperboloidal time, and~$r$ is a compactified radial coordinate along these slices such that it monotonically transitions from~$0$ to a finite positive constant~$r_\mathscr{I}$ as~$R$ goes from~$0$ to~$\infty$. In these coordinates, the metric becomes
\begin{align}
ds^2 = & -dt^2 - 2 H' R' dt \, dr + (R')^2 (1 - (H')^2) \, dr^2 + R^2 \, d \theta^2 + R^2 \, \sin^2 \theta \, d\phi^2 \, ,
\end{align}
giving the following outgoing and ingoing characteristic speeds
\begin{align}
c^r_{\pm} = \pm \frac{1}{R'\left(1\mp H'\right)} \, .
\end{align}
To ensure that these slices are hyperboloidal meeting $\mathscr{I}^+$, we set $c_+^r = C_+^R = 1$ everywhere~\cite{Zen11, Hil15, HilHarBug16}. This results in
\begin{align}
H' = 1 - \frac{1}{R'} \, , \quad \textrm{ and, thereby, } \quad c_-^r = - \frac{1}{2R'-1} \, .
\end{align}
Notice that~$c_-^r$ now monotonically varies from~$-1$ to $0$ from the origin to~$\mathscr{I}^+$. Taking~$H$ to be even, with $H(0) = 0$, we obtain
\begin{align}\label{eq:Height}
H(R(r)) = \left\{ \begin{array}{c}
 R(r) - r \, , \quad \textrm{for} \quad r \geq 0 \\
 r - R(r) \, , \quad \textrm{for} \quad r < 0 \, . 
\end{array} \right.
\end{align}
This function is~$C^\infty$ at the origin if~$R(r)$ satisfies
\begin{align}\label{eq:R_r}
R(0) = 0 \, , \, R'(0) = 1 \, , \textrm{ and } R^{(k)}(0) = 0 \quad \forall \, k \geq 2 \, .
\end{align}
Introducing the compactification as in~\cite{CalGunHil06},
\begin{align}\label{eq:Compactification}
& R(r) = \frac{r}{\Omega^{\frac{1}{n-1}}(r)} \, , \textrm{ with } r \in [0,r_\mathscr{I}] \, , \textrm{ and } 1 < n \leq 2 \, , \non \\
& \Omega(r) > 0 \textrm{ and } \Omega'(r) \leq 0 \textrm{ for } 0 < r < r_\mathscr{I} \, , \textrm{ and} \quad \Omega(r_\mathscr{I}) = 0 \, ,
\end{align}
the height function~\eqref{eq:Height} now couples the choice of compactification to that of slicing,  with the parameter~$n$ determining its asymptotic behavior, as~$R' \sim 2 R^n/(n-1)$ for large~$R$. The conditions~\eqref{eq:R_r} at the origin now translate to
\begin{align}\label{eq:Omega_Origin}
\Omega(0) = 1 \textrm{ and } \Omega^{(k)}(0) = 0 \quad \forall \, k > 0 \, .
\end{align}
One choice of~$\Omega$ satisfying all these requirements is
\begin{equation}\label{eq:Omega}
\Omega(r) = 1 - \frac{1}{2}\frac{r^2}{r_{\mathscr{I}}^2}\biggl[\tanh \biggl\{\tan \left(\pi\left(\frac{r}{r_{\mathscr{I}}} - \frac{1}{2}\right)\right)\biggl\}+1\biggl] \, .
\end{equation}
We will take~$r_\mathscr{I} = 1$ and~$n=2$, unless stated otherwise. The metric, in terms of these coordinates, now becomes
\begin{align}\label{eq:metric_hyp}
g_{\mu \nu} = \begin{pmatrix}
-1 & 1-R' & 0 & 0\\
 1-R' & 2R'-1 & 0 & 0\\
0 & 0 & R^2 & 0\\
0 & 0 & 0 & R^2\sin^2\theta
\end{pmatrix},
\end{align}
with the lapse $\alpha$, and shift $\beta^i$, given by
\begin{align}\label{eq:lapse_shift}
\alpha = \frac{R'}{\left(2R'-1\right)^{1/2}} \, , \quad \beta^\mu = \bigg\{ 0, -\frac{R'-1}{2R'-1}, 0, 0 \bigg\} \, .
\end{align}
The unit normal~$n^a$, spatial metric~$\gamma_{ab}$ and extrinsic curvature~$K_{ab}$ on these slices can be computed using the standard definitions.

\subsection{Linear Wave Equation}\label{sec:LWE}

Consider the following linear wave equation~(LWE):
\begin{equation}\label{eq:LWEP}
\left(\Box - F\right)\psi = 0,
\end{equation}
where~$\Box$ represents the standard d'Alembertian, the wave operator, on a Minkowski background, and~$F$ is a potential. For our current discussion, we will focus exclusively on those potentials~$F$ that are defined everywhere and are functions of spatial coordinates. In spherical polar coordinates, this wave equation is expressed as
\begin{align}\label{eq:LWEP_Sph_Coord}
& \bigg[-\p_T^2 + \frac{1}{R^2} \, \p_R \left( R^2 \, \p_R \right) + \frac{1}{R^2 \sin \theta} \, \p_\theta \left( \sin\theta \, \p_\theta \right) + \frac{1}{R^2 \sin^2 \theta} \, \p_\phi^2 - F \bigg] \psi = 0 \, ,
\end{align}
everywhere,
\begin{align}\label{eq:LWEP_Sph_Coord_z-axis_m=0}
\bigg[ - \p_T^2 + \frac{1}{R^2} \, \p_R \left( R^2 \, \p_R \right) + \frac{2}{R^2} \, \p_\theta^2 - F_{(m=0)} \bigg] \psi_{(m=0)} = 0 \, ,
\end{align}
on the~$z$-axis, and
\begin{align}\label{eq:LWEP_Sph_Coord_origin_l=0}
\left[ - \, \p_T^2 + 3 \, \p_R^2 - F_{(l=0)} \right] \psi_{(l=0)} = 0 \, ,
\end{align}
at the origin~\cite{GauRedKum26}. Here,~$\psi_{(m=0)}$ contains all the~$m=0$ modes in the spherical expansion of~$\psi$,
\begin{align}\label{eq:Psi_Sph_Harmonics_m=0}
\psi_{(m=0)} = \psi_{(m=0)}(T,R,\theta) \equiv \sum_l \psi_{l,0}(T,R) Y_{l,0}(\theta) \, ,
\end{align}
and,~$\psi_{(l=0)}$ contains only~$l=0$ mode,
\begin{align}\label{eq:Psi_Sph_Harmonics_l=0}
\psi_{(l=0)} = \psi_{(l=0)}(T,R) \equiv \psi_{0,0}(T,R) Y_{0,0} \, .
\end{align}
These modes can be extracted from the data if it is defined at all angular coordinates, from the following relations
\begin{align}\label{eq:m=0_mode}
\psi_{(m=0)}(T, R, \theta) = \frac{1}{2\pi} \int_0^{2\pi} \psi(T, R, \theta, \phi) \, d\phi \, ,
\end{align}
and
\begin{align}\label{eq:l=0_mode}
\psi_{(l=0)}(T, R) = \frac{1}{4\pi} \int_{\theta = 0}^{\pi} \int_{\phi = 0}^{2\pi} \psi(T, R, \theta, \phi) \, \sin\theta \, d\theta \, d\phi \, .
\end{align}
The functions~$F_{(m=0)}$ and~$F_{(l=0)}$ are defined similarly.

\subsection{First-Order Reduction}\label{sec:FOR}

One way to numerically evolve a second-order hyperbolic system is to introduce an equivalent first-order reduction~(FOR) system. In terms of our coordinates, a natural~FOR is
\begin{align}\label{eq:FOR_TR_comps}
& \psi_T \equiv \p_{T} \psi \, , \; \psi_R \equiv \p_{R} \psi \, , \; \psi_{\theta} \equiv \p_{\theta} \psi \, , \; \psi_{\phi} \equiv \p_{\phi} \psi \, ,
\end{align}
which we shall refer to as the Cauchy~FOR variables. It leads to the following~FOR system
\begin{align}\label{eq:LWEP_FOR_Constraint_damping}
\p_T \psi = &\ \psi_T \, , \quad
\p_T \psi_T = \frac{1}{R^2} \p_R \left( R^2 \, \psi_R \right)
 + \frac{1}{R^2 \sin \theta} \p_\theta \left( \sin\theta \: \psi_\theta \right) + \frac{1}{R^2 \sin^2 \theta} \p_\phi \psi_\phi - F \psi \, , \non \\
\p_T \psi_R = &\ \p_R \psi_T + \zeta_R (\p_R \psi - \psi_R) \, , \quad
\p_T \psi_\theta = \p_\theta \psi_T + \zeta_\theta (\p_\theta \psi -  \psi_\theta) \, , \non \\
\p_T \psi_\phi = &\ \p_\phi \psi_T + \zeta_\phi (\p_\phi \psi -  \psi_\phi) \, ,
\end{align}
where~$\mathcal{C}_{i'} \equiv \psi_{i'} - \p_{i'} \psi$ are the corresponding~FOR constraints, and the parameters~$\zeta_R$,~$\zeta_\theta$, and~$\zeta_\phi$ determine the rate of damping of these constraint violations, giving the following evolutions
\begin{align}
\hspace*{-1.0em} C_R (T) = C_{R} (0) \, e^{-\zeta_R T} \, , \quad  C_\theta (T) = C_{\theta} (0) \, e^{-\zeta_\theta T} \, , \quad
\textrm{and} \quad C_\phi (T) = C_{\phi} (0) \, e^{-\zeta_\phi T} \, .
\end{align}
As we observed in~\cite{GauRedKum26}, the resulting system is symmetric hyperbolic iff~$\zeta_R = \zeta_\theta = \zeta_\phi = \zeta$. With this choice, the above~FOR system reduces to
\begin{align}\label{eq:LWEP_FOR_Constraint_damping_z-axis}
\p_T \psi = &\ \psi_T \, , \quad
\p_T \psi_T = \frac{1}{R^2} \p_R \left( R^2 \, \psi_{R(m=0)} \right)
 + \frac{2}{R^2} \p_\theta \psi_{\theta(m=0)} - F_{(m=0)} \psi_{(m=0)} \, , \non \\
\p_T \psi_R = &\ \p_R \psi_T + \zeta \, (\p_R \psi - \psi_R) \, , \quad
\p_T \psi_\theta = \p_\theta \psi_T + \zeta \, (\p_\theta \psi -  \psi_\theta) \, , \quad
\p_T \psi_\phi = 0 \, ,
\end{align}
on the~$z$-axis, and to
\begin{align}\label{eq:LWEP_FOR_Constraint_damping_origin}
\p_T \psi = &\ \psi_T \, , \quad
\p_T \psi_T = 3 \, \p_R \psi_{R(l=0)} - F_{(l=0)} \psi_{(l=0)} \, , \non \\
\p_T \psi_R = &\ \p_R \psi_T + \zeta \, (\p_R \psi - \psi_R) \, , \quad
\p_T \psi_\theta = 0 \, , \quad
\p_T \psi_\phi = 0 \, ,
\end{align}
at the origin. Here~$\psi_{i'(m=0)}$ and~$\psi_{R(l=0)}$ are defined in the similar manner as~$\psi_{(m=0)}$ and~$\psi_{(l=0)}$ above.

In general,~$\psi$ satisfies the following parity conditions on the~$z$-axis and at the origin
\begin{align}\label{eq:Parity}
& \psi(T, R, -\theta, \phi) = \left\{
\begin{array}{cc}
\psi(T, R, \theta, \phi + \pi) \, , & \textrm{for} \, 0 \leq \phi < \pi \\
\psi(T, R, \theta, \phi - \pi) \, , & \textrm{for} \, \pi \leq \phi < 2\pi \, ,
\end{array}
\right. \non \\
& \psi(T, R, \pi + \theta, \phi) = \left\{
\begin{array}{c}
\psi(T, R,\pi - \theta, \phi + \pi) \, , \\
\textrm{for} \, 0 \leq \phi < \pi \\
\psi(T, R,\pi - \theta, \phi - \pi) \, , \\
\textrm{for} \, \pi \leq \phi < 2\pi \, ,
\end{array}
\right. \non \\
& \psi(T,-R,\theta,\phi) = \left\{
\begin{array}{c}
\psi(T, R, \pi - \theta, \phi + \pi) \, , \\
\textrm{for} \, 0 \leq \phi < \pi \\
\psi(T, R, \pi - \theta, \phi - \pi) \, , \\
\textrm{for} \, \pi \leq \phi < 2\pi \, .
\end{array}
\right. \, ,
\end{align}
and the~FOR variables satisfy
\begin{align}\label{eq:Parity_FOR_R_theta}
& \psi_\theta (T, R, -\theta, \phi) = \left\{
\begin{array}{cc}
- \psi_\theta (T, R, \theta, \phi + \pi) \, , & \textrm{for} \, 0 \leq \phi < \pi \\
- \psi_\theta (T, R, \theta, \phi - \pi) \, , & \textrm{for} \, \pi \leq \phi < 2\pi \, ,
\end{array}
\right. \non \\
& \psi_\theta (T, R, \pi + \theta, \phi) = \left\{
\begin{array}{c} 
- \psi_\theta (T, R,\pi - \theta, \phi + \pi) \, , \\
\textrm{for} \, 0 \leq \phi < \pi \\
- \psi_\theta (T, R,\pi - \theta, \phi - \pi) \, , \\
\textrm{for} \, \pi \leq \phi < 2\pi \, ,
\end{array}
\right. \non \\
& \psi_R (T,-R,\theta,\phi) = \left\{
\begin{array}{c}
- \psi_R (T, R, \pi - \theta, \phi + \pi) \, , \\
\textrm{for} \, 0 \leq \phi < \pi \\
- \psi_R (T, R, \pi - \theta, \phi - \pi) \, , \\
\textrm{for} \, \pi \leq \phi < 2\pi \, ,
\end{array}
\right. \, .
\end{align}
The parity conditions for the remaining~FOR variables remain the same as of~$\psi$ above. These conditions will be used to fill the ghost points outside the domain of~$\theta$ and for negative~$R$ during discretization. Additionally, we will use the periodicity condition to populate the ghost points beyond the domain of~$\phi$
\begin{align}\label{eq:Parity_phi}
& \psi(T, R, \theta, - \phi) = \psi(T, R, \theta, 2\pi - \phi) \, , \quad \psi(T, R, \theta, 2\pi + \phi) = \psi(T, R, \theta, \phi) \, ,
\end{align}
and the~$\phi$-derivatives do not change these conditions.

\subsection{Characteristic Variables}\label{sec:FOR_Characteristic}

As we observed in~\cite{GauRedKum26}, the system is entirely regular at~$\mathscr{I}^+$ when described using the characteristic~FOR variables
\begin{align}\label{eq:FOR_null_comps}
\psi_+ \equiv \psi_T + \psi_R \, , \quad \textrm{and,} \quad \psi_- \equiv \psi_T - \psi_R \, .
\end{align}
This leads to the following~FOR system
\begin{align}\label{eq:LWEP_Characteristic_FOR_Constraint_damping}
\p_T \psi = &\ \frac{\psi_+ + \psi_-}{2} \, , \non \\
\p_T \psi_+ = &\ \p_R \psi_+ + \frac{\psi_+ - \psi_-}{R} + \frac{1}{R^2 \sin \theta} \p_\theta \left( \sin\theta \: \psi_\theta \right) + \frac{1}{R^2 \sin^2 \theta} \p_\phi \psi_\phi - F \psi \non \\ 
& + \zeta \left( \p_R \psi - \frac{\psi_+ - \psi_-}{2} \right) \, , \non \\
\p_T \psi_- = &\ - \p_R \psi_- + \frac{\psi_+ - \psi_-}{R} + \frac{1}{R^2 \sin \theta} \p_\theta \left( \sin\theta \: \psi_\theta \right) + \frac{1}{R^2 \sin^2 \theta} \p_\phi \psi_\phi - F \psi \non \\ 
& - \zeta \left( \p_R \psi - \frac{\psi_+ - \psi_-}{2} \right) \, , \non \\
\p_T \psi_\theta = &\ \frac{\p_\theta (\psi_+ + \psi_-)}{2} + \zeta \, (\p_\theta \psi -  \psi_\theta) \, , \quad
\p_T \psi_\phi = \frac{\p_\phi (\psi_+ + \psi_-)}{2} + \zeta \, (\p_\phi \psi -  \psi_\phi) \, ,
\end{align}
everywhere, 
\begin{align}\label{eq:LWEP_Characteristic_FOR_Constraint_damping_z-axis}
\p_T \psi = &\ \frac{\psi_+ + \psi_-}{2} \, , \non \\
\p_T \psi_+ = &\ \p_R \psi_+ + \frac{\psi_+ - \psi_-}{R} + \frac{2}{R^2} \p_\theta \psi_{\theta(m=0)} - F_{(m=0)} \psi_{(m=0)} + \zeta \left( \p_R \psi - \frac{\psi_+ - \psi_-}{2} \right) \, , \non \\
\p_T \psi_- = &\ - \p_R \psi_- + \frac{\psi_+ - \psi_-}{R} + \frac{2}{R^2} \p_\theta \psi_{\theta(m=0)} - F_{(m=0)} \psi_{(m=0)} - \zeta \left( \p_R \psi - \frac{\psi_+ - \psi_-}{2} \right) \, , \non \\
\p_T \psi_\theta = &\ \frac{\p_\theta (\psi_+ + \psi_-)}{2} + \zeta \, (\p_\theta \psi -  \psi_\theta) \, , \quad
\p_T \psi_\phi = 0 \, .
\end{align}
on the~$z$-axis, and
\begin{align}\label{eq:LWEP_Characteristic_FOR_Constraint_damping_origin}
\p_T \psi = &\ \frac{\psi_+ + \psi_-}{2} \, , \non \\
\p_T \psi_+ = &\ \frac{3 \, \p_R (\psi_+ - \psi_-)_{(l=0)}}{2} + \frac{\p_R (\psi_+ + \psi_-)}{2} - F_{(l=0)} \psi_{(l=0)} + \zeta \left( \p_R \psi - \frac{\psi_+ - \psi_-}{2} \right) \, , \non \\
\p_T \psi_- = &\ \frac{3 \, \p_R (\psi_+ - \psi_-)_{(l=0)}}{2} - \frac{\p_R (\psi_+ + \psi_-)}{2} - F_{(l=0)} \psi_{(l=0)} - \zeta \left( \p_R \psi - \frac{\psi_+ - \psi_-}{2} \right) \, , \non \\
\p_T \psi_\theta = &\ 0 \, , \quad
\p_T \psi_\phi = 0 \, ,
\end{align}
at the origin. Here, we do not extract the~$m=0$ and $l=0$ modes in all terms on the right, as not all arise from the~$\psi_T$ equation in~\eqref{eq:LWEP_FOR_Constraint_damping_z-axis} and \eqref{eq:LWEP_FOR_Constraint_damping_origin}. The parity conditions for these~FOR variables become
\begin{align}\label{eq:Parity_FOR_Null_R}
\psi_\pm (T,-R,\theta,\phi) = \left\{
\begin{array}{c}
\psi_\mp (T, R, \pi - \theta, \phi + \pi) \, , \\
\textrm{for} \, 0 \leq \phi < \pi \\
\psi_\mp (T, R, \pi - \theta, \phi - \pi) \, , \\
\textrm{for} \, \pi \leq \phi < 2\pi \, ,
\end{array}
\right.
\end{align}
keeping the remaining ones unchanged. Due to these parity conditions, $\p_R (\psi_+ - \psi_-)_{(l=0)} = 2 \, \p_R (\psi_+)_{(l=0)} = - 2 \, \p_R (\psi_-)_{(l=0)}$ at the origin, so the~$\psi_+$ and~$\psi_-$ equations in~\eqref{eq:LWEP_Characteristic_FOR_Constraint_damping_origin} could be rewritten accordingly.

\subsection{Introducing Hyperboloidal Coordinates: Dual Foliation approach}\label{sec:LWEP_Hyp_Slices}

We now express these equations using hyperboloidal coordinates, employing the dual-foliation formulation proposed in~\cite{Hil15, HilHarBug16}, which we employed in our previous works~\cite{GasGauHil20, GauVanHil21, PetGauRai23, PetGauVan24, PetHil25}. This formulation uses the~$3+1$ split of the Jacobian~$J^{\mu'}_\mu = \p X^{\mu'}/ \p x^\mu$ which relates the coordinate systems~$X^{\mu'} = (T,R, \theta, \phi)$ and~$x^\mu = (t, r, \theta, \phi)$, and, thereby, decouples the choice of gauge from the choice of coordinates. As a result, one can use the same~FOR variables, defined with respect to the coordinates~$X^{\nu'}$, as functions of the coordinates~$x^\mu$. In terms of the hyperboloidal coordinates described in Sec.~\ref{sec:Hyperboloidal_slices}, the time and radial derivatives transform as
\begin{align}
\p_T = \p_t \, , \quad \textrm{and} \quad \p_R = \frac{(R'-1) \, \p_t + \p_r}{R'} \, ,
\end{align}
giving rise to the following~FOR system
\begin{align}\label{eq:hyp_LWEP_FOR_null}
\p_t \psi = &\ \frac{1}{2} \left( \psi_+ + \psi_- \right) \, , \non \\
\p_t \psi_+ = &\ \frac{1}{2R'-1} \bigg[ \p_r \psi_+ + \frac{R'}{R} (\psi_+ - \psi_-) + \frac{R'}{R^2} \bigg( \p_\theta \psi_\theta + \cot\theta \, \psi_\theta + \frac{1}{\sin^2\theta} \p_\phi \psi_\phi \bigg) - R' F \psi \non \\
& + \zeta \, \left( \p_r \psi - \frac{(2R'-1) \psi_+}{2} + \frac{\psi_-}{2} \right) \bigg] \, , \non \\
\p_t \psi_- = &\ - \p_r \psi_- + \frac{R'}{R} (\psi_+ - \psi_-) + \frac{R'}{R^2} \bigg( \p_\theta \psi_\theta + \cot\theta \, \psi_\theta + \frac{1}{\sin^2\theta} \p_\phi \psi_\phi \bigg) - R' F \psi \non \\
& - \zeta \, \left( \p_r \psi - \frac{(2R'-1) \psi_+}{2} + \frac{\psi_-}{2} \right)  \, , \non \\
\p_t \psi_\theta = &\ \frac{1}{2} \p_\theta (\psi_+ + \psi_-) + \zeta \left( \p_\theta \psi -\psi_\theta \right) \, , \quad
\p_t \psi_\phi = \frac{1}{2} \p_\phi (\psi_+ + \psi_-) + \zeta \left( \p_\phi \psi -\psi_\phi \right) \, .
\end{align}
The equations at the poles and the origin can be written similarly. Notably, the equations at the origin remain the same whenever the conditions~\eqref{eq:R_r} are satisfied.

\subsection{Regularisation at~$\mathscr{I}^+$}
\label{sec:Reg_LWE_Hyp_Slices}

In Sec.~\ref{sec:Hyperboloidal_slices}, it was noted that~$R' \sim 2 R^n/(n-1)$ for large~$R$, with~$1 < n \leq 2$, making many terms in eq.~\eqref{eq:hyp_LWEP_FOR_null} singular at~$\mathscr{I}^+$. However, these terms are only formally singular as the solution to the linear wave equation falls off like~$1/R$ towards~$\mathscr{I}^+$, and any positive~$F$ makes it fall-off faster. Therefore, we regularize these equations at~$\mathscr{I}^+$ by rescaling the variables as
\begin{align}\label{eq:Rescaled_FOR_null_comps}
& \tilde{\psi} \equiv \chi \, \psi \, , \quad \tilde{\psi}_+ \equiv \chi_+ \psi_+ \, , \quad \tilde{\psi}_- \equiv \chi_- \psi_- \, , \quad \tilde{\psi}_\theta \equiv \chi \, \psi_\theta \, , \quad \tilde{\psi}_\phi \equiv \chi \, \psi_\phi \, , \non \\
\textrm{with} \quad & \chi_+ = \chi^2 \quad \textrm{and} \quad \chi_- = \chi \quad \textrm{for} \quad r > 0 \, .
\end{align}
Here,~$\chi = \chi(r)$ is monotonically increasing, with~$\chi(0) = 1$ and~$\sim R$ for large~$R$. These rescalings stem from the asymptotic behaviour of the characteristic variables towards~$\mathscr{I}^+$ as described in~\cite{GasHil19}. The parity of these variables are kept the same as those of the unrescaled ones~\eqref{eq:Parity_FOR_Null_R} by imposing the following conditions on~$\chi$ and~$\chi_\pm$
\begin{align}\label{eq:Parity_Chi}
\chi(-r) = \chi(r) \, , \quad \chi_\pm(-r) = \chi_\mp (r) \, .
\end{align}
For smoothness, we choose~$\chi$ such that~$\chi$ and~$\chi_\pm$ are~$C^\infty$ at the origin. As noetd in~\cite{GauRedKum26}, the most natural choice for~$\chi$ is
\begin{align}\label{eq:Chi_natural}
\chi = \Omega^{\frac{1}{1-n}} \, ,
\end{align}
which corresponds to conformal rescaling in the case of conformal compactification as discussed in~\cite{Zen07, Zen08, ZenKid10, Zen11, ZenGal12, YanZimZen13, Rin25, MonRin08, Rin09, Rin10, RinMon13, Rin14, BaaRin16, MalRin18, VanHusHil14, VanHus14, Van15, VanHus16, VanHus17, Van23, Van23a, VanVal24, AlvVan25, AlvVan25a}, for~$n=2$, and satisfies all the required properties outlined above for~$\Omega$ satisfying~\eqref{eq:Omega_Origin}. Additionally, it gives~$R/\chi = r$, which simplifies the final equations and the volume element presented in~\eqref{eq:Energy_Density_Rescaled_Variables}. Consequently, this leads to the following~FOR system
\begin{align}\label{eq:hyp_LWEP_rescaled_FOR_null}
\p_t \tilde{\psi} = &\ \frac{1}{2} \left( \frac{\tilde{\psi}_+}{\chi} + \tilde{\psi}_- \right) \, , \non  \\
\p_t \tilde{\psi}_+ = &\ \frac{1}{2R'-1} \bigg[ \bigg( \p_r + \frac{R'}{R} - \frac{2 \chi'}{\chi} \bigg) \tilde{\psi}_+ - \frac{\chi R'}{R} \tilde{\psi}_- + \frac{\chi R'}{R^2} \bigg( \p_\theta \tilde{\psi}_\theta + \cot\theta \, \tilde{\psi}_\theta + \frac{1}{\sin^2\theta} \p_\phi \tilde{\psi}_\phi \bigg) \non \\
& - \chi R' F \tilde{\psi} + \zeta \, \bigg( \chi \p_r \tilde{\psi} - \chi' \tilde{\psi} - \frac{(2R'-1) \tilde{\psi}_+}{2} + \frac{\chi \tilde{\psi}_-}{2} \bigg) \bigg] \, , \non \\
\p_t \tilde{\psi}_- = &\ \bigg[ - \bigg( \p_r + \frac{R'}{R} - \frac{\chi'}{\chi} \bigg) \tilde{\psi}_- + \frac{R'}{R \chi} \tilde{\psi}_+ + \frac{R'}{R^2} \bigg( \p_\theta \tilde{\psi}_\theta + \cot\theta \, \tilde{\psi}_\theta + \frac{1}{\sin^2\theta} \p_\phi \tilde{\psi}_\phi \bigg) - R' F \tilde{\psi} \non \\
& - \zeta \, \bigg( \p_r \tilde{\psi} - \frac{\chi'}{\chi} \tilde{\psi} - \frac{(2R'-1) \tilde{\psi}_+}{2 \chi} + \frac{\tilde{\psi}_-}{2} \bigg) \bigg] \, , \non \\
\p_t \tilde{\psi}_\theta = &\ \frac{1}{2} \p_\theta \left( \frac{\tilde{\psi}_+}{\chi} + \tilde{\psi}_- \right) + \zeta \left( \p_\theta \tilde{\psi} -\tilde{\psi}_\theta \right) \, , \quad
\p_t \tilde{\psi}_\phi = \frac{1}{2} \p_\phi \left( \frac{\tilde{\psi}_+}{\chi} + \tilde{\psi}_- \right) + \zeta \left( \p_\phi \tilde{\psi} -\tilde{\psi}_\phi \right) \, .
\end{align}
As~$\chi \sim R$ and~$\chi' \sim R'$ for large~$R$, these equations reduce to the following system at~$\mathscr{I}^+$ for~$n=2$, and taking~$\tilde{\zeta} \, \equiv \zeta \, \, \chi$:
\begin{align}\label{eq:hyp_LWEP_rescaled_FOR_null_scri}
\p_t \tilde{\psi} = &\ \frac{1}{2} \tilde{\psi}_- \, , \quad
\p_t \tilde{\psi}_+ = - \frac{1}{2} \tilde{\psi}_- - \frac{R F}{2} \tilde{\psi} \, , \non \\
\p_t \tilde{\psi}_- = &\ - \p_r \tilde{\psi}_- + 2 \, \tilde{\psi}_+ + 2 \, \bigg( \p_\theta \tilde{\psi}_\theta + \cot\theta \, \tilde{\psi}_\theta + \frac{1}{\sin^2\theta} \p_\phi \tilde{\psi}_\phi \bigg) - R' F \tilde{\psi} + 2 \, \tilde{\zeta} \, \, ( \tilde{\psi} + \tilde{\psi}_+ ) \, , \non \\
\p_t \tilde{\psi}_\theta = &\ \frac{1}{2} \p_\theta \tilde{\psi}_- + \frac{\tilde{\zeta}}{\chi} \left( \p_\theta \tilde{\psi} -\tilde{\psi}_\theta \right) \, , \quad
\p_t \tilde{\psi}_\phi = \frac{1}{2} \p_\phi \tilde{\psi}_- + \frac{\tilde{\zeta}}{\chi} \left( \p_\phi \tilde{\psi} -\tilde{\psi}_\phi \right) \, ,
\end{align}
This system is fully regular at~$\mathscr{I}^+$ if~$\tilde{\zeta} = O(1)$, and the potential~$F$ falls-off like~$1/R^2$ for large~$R$. For a fall-off like~$1/R^{1+\epsilon}$ in~$F$, with~$0 < \epsilon < 1$, the above system can be made regular by taking~$n \in (1, 1 + \epsilon]$, cf~\cite{GauVanHil21}. However, an even slower fall-off makes the system singular at~$\mathscr{I}^+$. One such example is~$F = m^2$. Here, we must impose a sufficient fall-off in the~ID to ensure the required decay of the solution towards~$\mathscr{I}^+$ for all times~\cite{Kla93, GauVanHil21}.

\subsection{Introducing Regularized Covariant Divergence Operators}
\label{sec:Reg_Cov_Div}

We next introduce the spatial divergence operator required to derive the~$3$D~SBP scheme. Its components along the three coordinate directions are as follows:
\begin{align}
\hspace*{-1.0em} \bigg( \p_r + \frac{2R'}{R} \bigg) f = \frac{1}{R^2} \p_r (R^2 \, f) \, , \, \frac{1}{\sin\theta} \p_\theta (\sin\theta \, f) \, , \, \p_\phi f \, .
\end{align}
However, as~$r \rightarrow r_\mathscr{I}$,~$R \rightarrow \infty$, and it is numerically infeasible to define the first operator at~$\mathscr{I}^+$. Therefore, we define the following regular operators
\begin{align}\label{eq:Tilded_Operators}
& \tilde{\p}_r f \equiv \frac{\chi^2}{R^2} \p_r \left( \frac{R^2}{\chi^2} f \right) = \p_r f + 2 \left( \frac{R'}{R} - \frac{\chi'}{\chi} \right) f \, , \non \\
& \tilde{\p}_\theta f \equiv \frac{1}{\sin\theta} \p_\theta \left( (\sin\theta) \, f \right) = \p_\theta f + (\cot\theta) \, f \, , \quad \textrm{and} \quad
 \tilde{\p}_\phi \equiv \p_\phi \, .
\end{align}
For~$\chi$ given in~\eqref{eq:Chi_natural}, the above definition simplifies~$\tilde{\p}_r$ to
\begin{align}\label{eq:Tilded_Operator_simplified}
\tilde{\p}_r f = r^{-2} \p_r (r^2 f) \, .
\end{align}

Substituting these operators in~\eqref{eq:hyp_LWEP_rescaled_FOR_null} gives
\begin{align}\label{eq:hyp_LWEP_rescaled_FOR_null_Tilded}
\p_t \tilde{\psi} = &\ \frac{1}{2} \left( \frac{\tilde{\psi}_+}{\chi} + \tilde{\psi}_- \right) \, , \non  \\
\p_t \tilde{\psi}_+ = &\ \frac{\chi}{2R'-1} \bigg[ \bigg( \frac{\p_r + \tilde{\p}_r}{2} \bigg) \left( \frac{\tilde{\psi}_+}{\chi} \right) + \bigg( \frac{\p_r - \tilde{\p}_r}{2} \bigg) \tilde{\psi}_- - \frac{\chi'}{\chi} \tilde{\psi}_- + \frac{R'}{R^2} \bigg( \tilde{\p}_\theta \tilde{\psi}_\theta + \frac{1}{\sin^2\theta} \p_\phi \tilde{\psi}_\phi \bigg) \non \\
& - R' F \tilde{\psi} + \frac{\tilde{\zeta}}{\chi} \, \bigg( \p_r \tilde{\psi} - \frac{\chi'}{\chi} \tilde{\psi} - \frac{(2R'-1) \tilde{\psi}_+}{2 \chi} + \frac{\tilde{\psi}_-}{2} \bigg) \bigg] \, , \non \\
\p_t \tilde{\psi}_- = &\ \bigg[ - \bigg( \frac{\p_r + \tilde{\p}_r}{2} \bigg) \tilde{\psi}_- - \bigg( \frac{\p_r - \tilde{\p}_r}{2} \bigg) \left( \frac{\tilde{\psi}_+}{\chi} \right) + \frac{\chi'}{\chi^2} \tilde{\psi}_+ + \frac{R'}{R^2} \bigg( \tilde{\p}_\theta \tilde{\psi}_\theta + \frac{1}{\sin^2\theta} \p_\phi \tilde{\psi}_\phi \bigg) \non \\
& - R' F \tilde{\psi} - \frac{\tilde{\zeta}}{\chi} \, \bigg( \p_r \tilde{\psi} - \frac{\chi'}{\chi} \tilde{\psi} - \frac{(2R'-1) \tilde{\psi}_+}{2 \chi} + \frac{\tilde{\psi}_-}{2} \bigg) \bigg] \, , \non \\
\p_t \tilde{\psi}_\theta = &\ \frac{1}{2} \p_\theta \left( \frac{\tilde{\psi}_+}{\chi} + \tilde{\psi}_- \right) + \frac{\tilde{\zeta}}{\chi} \left( \p_\theta \tilde{\psi} -\tilde{\psi}_\theta \right) \, , \quad
\p_t \tilde{\psi}_\phi = \frac{1}{2} \p_\phi \left( \frac{\tilde{\psi}_+}{\chi} + \tilde{\psi}_- \right) + \frac{\tilde{\zeta}}{\chi} \left( \p_\phi \tilde{\psi} -\tilde{\psi}_\phi \right) \, ,
\end{align}
everywhere,
\begin{align}\label{eq:hyp_LWEP_rescaled_FOR_null_Tilded_z-axis}
\p_t \tilde{\psi} = &\ \frac{1}{2} \left( \frac{\tilde{\psi}_+}{\chi} + \tilde{\psi}_- \right) \, , \non  \\
\p_t \tilde{\psi}_+ = &\ \frac{\chi}{2R'-1} \bigg[ \bigg( \frac{\p_r + \tilde{\p}_r}{2} \bigg) \left( \frac{\tilde{\psi}_+}{\chi} \right) + \bigg( \frac{\p_r - \tilde{\p}_r}{2} \bigg) \tilde{\psi}_- - \frac{\chi'}{\chi} \tilde{\psi}_- + \frac{2 R'}{R^2} \p_\theta \tilde{\psi}_{\theta (m=0)} - R' F \tilde{\psi} \non \\
& + \frac{\tilde{\zeta}}{\chi} \, \bigg( \p_r \tilde{\psi} - \frac{\chi'}{\chi} \tilde{\psi} - \frac{(2R'-1) \tilde{\psi}_+}{2 \chi} + \frac{\tilde{\psi}_-}{2} \bigg) \bigg] \, , \non \\
\p_t \tilde{\psi}_- = &\ \bigg[ - \bigg( \frac{\p_r + \tilde{\p}_r}{2} \bigg) \tilde{\psi}_- - \bigg( \frac{\p_r - \tilde{\p}_r}{2} \bigg) \left( \frac{\tilde{\psi}_+}{\chi} \right) + \frac{\chi'}{\chi^2} \tilde{\psi}_+ + \frac{2 R'}{R^2} \p_\theta \tilde{\psi}_{\theta (m=0)} - R' F \tilde{\psi} \non \\
& - \frac{\tilde{\zeta}}{\chi} \, \bigg( \p_r \tilde{\psi} - \frac{\chi'}{\chi} \tilde{\psi} - \frac{(2R'-1) \tilde{\psi}_+}{2 \chi} + \frac{\tilde{\psi}_-}{2} \bigg) \bigg] \, , \non \\
\p_t \tilde{\psi}_\theta = &\ \frac{1}{2} \p_\theta \left( \frac{\tilde{\psi}_+}{\chi} + \tilde{\psi}_- \right) + \frac{\tilde{\zeta}}{\chi} \left( \p_\theta \tilde{\psi} -\tilde{\psi}_\theta \right) \, , \quad
\p_t \tilde{\psi}_\phi = 0 \, ,
\end{align}
on the~$z$-axis, and
\begin{align}\label{eq:hyp_LWEP_rescaled_FOR_null_Tilded_origin}
\p_t \tilde{\psi} = &\ \frac{\tilde{\psi}_+ + \tilde{\psi}_-}{2} \, , \non  \\
\p_t \tilde{\psi}_+ = &\ \frac{3 \, \p_r (\tilde{\psi}_+ - \tilde{\psi}_-)_{(l=0)}}{2} + \frac{\p_r (\tilde{\psi}_+ + \tilde{\psi}_-)}{2} - F \tilde{\psi}_{(l=0)} + \tilde{\zeta} \, \bigg( \p_r \tilde{\psi} - \frac{\tilde{\psi}_+ - \tilde{\psi}_-}{2} \bigg) \, , \non \\
\p_t \tilde{\psi}_- = &\ \frac{3 \, \p_r (\tilde{\psi}_+ - \tilde{\psi}_-)_{(l=0)}}{2} - \frac{\p_r (\tilde{\psi}_+ + \tilde{\psi}_-)}{2} - F \tilde{\psi}_{(l=0)} - \tilde{\zeta} \, \bigg( \p_r \tilde{\psi} - \frac{\tilde{\psi}_+ - \tilde{\psi}_-}{2} \bigg) \, , \non \\
\p_t \tilde{\psi}_\theta = &\ 0 \, , \quad
\p_t \tilde{\psi}_\phi = 0 \, ,
\end{align}
at the origin. For~$n=2$, these equations reduce to the following system at~$\mathscr{I}^+$,
\begin{align}\label{eq:hyp_LWEP_rescaled_FOR_null_Tilded_scri}
\p_t \tilde{\psi} = &\ \frac{\tilde{\psi}_-}{2} \, , \quad
\p_t \tilde{\psi}_+ = - \frac{\tilde{\psi}_-}{2} - \frac{R F}{2} \tilde{\psi} \, , \non \\
\p_t \tilde{\psi}_- = &\ - \bigg( \frac{\p_r + \tilde{\p}_r}{2} \bigg) \tilde{\psi}_- - \bigg( \frac{\p_r - \tilde{\p}_r}{2} \bigg) \left( \frac{\tilde{\psi}_+}{\chi} \right) + 2 \, \tilde{\psi}_+ + 2 \bigg( \tilde{\p}_\theta \tilde{\psi}_\theta + \frac{1}{\sin^2\theta} \p_\phi \tilde{\psi}_\phi \bigg) \non \\
& - R' F \, \tilde{\psi} + 2 \, \tilde{\zeta} \, ( \tilde{\psi} + \tilde{\psi}_+ ) \, , \non \\
\p_t \tilde{\psi}_\theta = &\ \frac{1}{2} \, \p_\theta \tilde{\psi}_- \, , \quad
\p_t \tilde{\psi}_\phi = \frac{1}{2} \, \p_\phi \tilde{\psi}_- \, ,
\end{align}
everywhere, and
\begin{align}\label{eq:hyp_LWEP_rescaled_FOR_null_Tilded_z-axis_scri}
\p_t \tilde{\psi} = &\ \frac{\tilde{\psi}_-}{2} \, , \quad
\p_t \tilde{\psi}_+ = - \frac{\tilde{\psi}_-}{2} - \frac{R F}{2} \tilde{\psi} \, , \non \\
\p_t \tilde{\psi}_- = &\ - \bigg( \frac{\p_r + \tilde{\p}_r}{2} \bigg) \tilde{\psi}_- - \bigg( \frac{\p_r - \tilde{\p}_r}{2} \bigg) \left( \frac{\tilde{\psi}_+}{\chi} \right) + 2 \, \tilde{\psi}_+ + 4 \, \p_\theta \tilde{\psi}_{\theta (m=0)} - R' F \, \tilde{\psi} \non \\
& + 2 \, \tilde{\zeta} \, ( \tilde{\psi} + \tilde{\psi}_+ ) \, , \non \\
\p_t \tilde{\psi}_\theta = &\ \frac{1}{2} \p_\theta \tilde{\psi}_- \, , \quad
\p_t \tilde{\psi}_\phi = 0 \, ,
\end{align}
on the~$z$-axis. These are the final equation used to derive the~SBP scheme. We will take~$\tilde{\zeta} = 0$ or~$1$, for all our purposes.

\subsection{Conserved energy on hyperboloidal slices}
\label{sec:Cons_Energy_hyp_slices}

As the~SBP scheme preserves conserved energy at the discrete level, the next and final step is to derive it on these slices. In terms of the rescaled characteristic~FOR variables, this energy is given by
\begin{align}\label{eq:Energy_norm}
E(t) \equiv \int_{\Sigma_t} \varepsilon(t,r,\theta,\phi) \, dr \, d\theta \, d\phi \, ,
\end{align}
with
\begin{align}\label{eq:Energy_Density_Rescaled_Variables}
\varepsilon(t,r,\theta,\phi) = &\ \frac{1}{2} \Bigg( F R' \, \tilde{\psi}^2 + \left( \frac{2R'-1}{2 \chi^2} \right) \tilde{\psi}_+^2 + \frac{\tilde{\psi}_-^2}{2} + \frac{R'}{R^2} \bigg( \tilde{\psi}_\theta^2 + \frac{1}{\sin^2 \theta} \tilde{\psi}_\phi^2  \bigg) \Bigg) \frac{R^2}{\chi^2} \, \sin\theta \, .
\end{align}

To calculate the flux at~$\mathscr{I}^+$, we first truncate these slices at~$r_o < r_\mathscr{I}$, and set~$r_o = r_o(t)$. The total energy on these truncated slices can now be defined as
\begin{align}
E(t,r_o (t)) \equiv \int_{r=0}^{r_o(t)} \int_{\theta=0}^\pi \int_{\phi=0}^{2\pi} \varepsilon(t,r,\theta,\phi) \, dr \, d\theta \, d\phi \, ,
\end{align}
giving
\begin{align}\label{eq:Energy_change}
\frac{d}{dt} E(t,r_o (t)) = \p_t E(t,r_o (t)) + \p_{r_o} E(t,r_o (t)) \frac{d r_o}{dt} \, .
\end{align}
Substituting the~EOMs, with~$\tilde{\zeta} = 0$, and using Gauss's theorem on the spatial slices, the first term on the right gives
\begin{align}\label{eq:E-dot_rescaled}
\p_t E(t, & r_o (t)) \equiv \int_{r=0}^{r_o(t)} \int_{\theta=0}^\pi \int_{\phi=0}^{2\pi} \p_t \varepsilon \, dr \, d\theta \, d\phi = \frac{1}{4} \int_{\theta=0}^\pi \int_{\phi=0}^{2\pi} \left( \frac{\tilde{\psi}_+^2}{\chi^2} - \tilde{\psi}_-^2 \right) \, \frac{R^2}{\chi^2} \, \sin\theta \, d\theta \, d\phi \, \bigg|_{r = r_o} \, ,
\end{align}
while the definition of~$E(t,r_o (t))$ gives
\begin{align}\label{eq:Moving_Boundary}
& \p_{r_o} E(t,r_o (t)) = \lim_{\delta r_o \rightarrow 0} \frac{1}{\delta r_o} \left( \int_{r=0}^{r_o + \delta r_o} - \int_{r=0}^{r_o(t)} \right) \int_{\theta=0}^\pi \int_{\phi=0}^{2\pi} \varepsilon \, dr \, d\theta \, d\phi = \int_{\theta=0}^\pi \int_{\phi=0}^{2\pi} \varepsilon \, d\theta \, d\phi \, \bigg|_{r = r_o} \, .
\end{align}
The outer boundary becomes an incoming-null hypersurface if we take~$dr_o/dt = c_-^r|_{r = r_o} = -1/(2R'(r_o) - 1)$, giving
\begin{align}\label{eq:Energy_change_expanded_rescaled}
\frac{d}{dt} E(t,r_o (t)) & = - \int_{\theta=0}^\pi \int_{\phi=0}^{2\pi} \frac{1}{2} \Bigg( F \, \tilde{\psi}^2 + \tilde{\psi}_-^2 + \frac{1}{R^2} \bigg( \tilde{\psi}_\theta^2 + \frac{1}{\sin^2 \theta} \tilde{\psi}_\phi^2 \bigg) \Bigg) \, \frac{R^2}{\chi^2} \, \frac{R'}{2R' - 1} \, \sin\theta \, d\theta \, d\phi \, \bigg|_{r = r_o} .
\end{align}
The total change in energy can now be determined by taking the limit~$r_o \rightarrow r_\mathscr{I}$, giving
\begin{align}\label{eq:Stokes_diff_form_rescaled}
\frac{d}{dt} E(t) = &\ - \int_{\theta=0}^\pi \int_{\phi=0}^{2\pi} \frac{1}{4} \left( F \, \tilde{\psi}^2 + \tilde{\psi}_-^2 \right) \, \sin\theta \, d\theta \, d\phi \, \bigg|_{r = r_\mathscr{I}} \, ,
\end{align}
which has a much simpler form and does not contain contributions from the angular variables~$\tilde{\psi}_\theta$ and~$\tilde{\psi}_\phi$, as these contributions fall off like~$1/R^2$.

\section{Discretization}\label{sec:Discretization}

We derive here a semi-discrete approximation of the continuum equations that discretizes only spatial coordinates, keeping the continuity in time intact. The spatial coordinates take discrete values
\begin{align}
& r_I \equiv I \Delta r \, , \; \theta_J \equiv J \Delta \theta \, , \; \phi_K \equiv K \Delta \phi \, , \quad
\textrm{with } \Delta r \equiv r_\mathscr{I}/N_r \, , \, \Delta \theta \equiv \pi/N_\theta \, , \, \Delta \phi \equiv 2\pi/N_\phi \, , \non \\
& I = 0, 1, 2, \ldots, N_r \, , \quad J = 0, 1, 2, \ldots, N_\theta \, , \quad K = 0, 1, 2, \ldots, N_\phi - 1 \, ,
\end{align}
and~$r_\mathscr{I} = 1$. All the variables are now defined as arrays of these integers and are denoted in uppercase. For example, we define~$(\Psi_\alpha)_{IJK} \equiv \psi_\alpha(r_I, \theta_J, \phi_K)$, with~$\psi_\alpha$ denoting any of the variables~$\psi$, $\psi_+$, $\psi_-$, $\psi_\theta$, or~$\psi_\phi$. In general, we define the state vector~$\mathbf{U}_{I J K} \equiv \mathbf{u}(r_I, \theta_J, \phi_K)$, with~$\mathbf{U} = \{ U_\alpha \} \equiv (\tilde{\Psi}, \tilde{\Psi}_+, \tilde{\Psi}_-, \tilde{\Psi}_\theta, \tilde{\Psi}_\phi)^T$.

Furthermore,~$\Upsilon$ denotes the quadrature matrix on the grid space, a discrete version of~$dr \, d\theta \, d\phi$, and encodes the information about the grid spacing at the point~$(I,J,K)$, and can be chosen to decompose as
\begin{align}\label{eq:Quadrature}
& \Upsilon_{I'J'K'IJK} \equiv \Upsilon^r_{I'I} \, \Upsilon^\theta_{J'J} \, \Upsilon^\phi_{K'K} \, , \quad
\textrm{with} \quad \Upsilon^r = {\rm diag}(\Delta r/2, \Delta r, \ldots, \Delta r, \Delta r/2) \, , \non \\
& \Upsilon^\theta = {\rm diag}(\Delta\theta/2, \Delta\theta, \ldots, \Delta\theta, \Delta\theta/2) \, , \quad \Upsilon^\phi = {\rm diag}(\Delta\phi, \Delta\phi, \ldots, \Delta\phi, \Delta\phi) \, .
\end{align}

We denote the discrete versions of $\p_r$, $\p_\theta$ and $\p_\phi$ with $(\Upsilon^r)^{-1} D_r$, $(\Upsilon^\theta)^{-1} D_\theta$ and $(\Upsilon^\phi)^{-1} D_\phi$, and of $\tilde{\p}_r$, $\tilde{\p}_\theta$ and $\tilde{\p}_\phi$ with $(\Upsilon^r)^{-1} \tilde{D}_r$, $(\Upsilon^\theta)^{-1} \tilde{D}_\theta$ and $(\Upsilon^\phi)^{-1} \tilde{D}_\phi$, respectively, to keep the discrete operators $D_r$, $D_\theta$, $D_\phi$, $\tilde{D}_r$, $\tilde{D}_\theta$ and $\tilde{D}_\phi$, dimensionless, and the time derivative with a dot overhead. Similarly, we take all the multiplicative operators $[f(r,\theta,\phi)]$ to be diagonal matrices with components
\begin{align}\label{eq:Discrete_Coefficients}
\hspace*{-1.0em} [f(r,\theta,\phi)]_{I J K I' J' K'} = \delta^r_{I I'} \, \delta^\theta_{J J'} \, \delta^\phi_{K K'} \, f(r_I, \theta_J, \phi_K) \, ,
\end{align}
with~$\delta^r_{II'}$,~$\delta^\theta_{JJ'}$ and~$\delta^\phi_{KK'}$ denoting Kronecker deltas of obvious dimensionalities. This way, all the multiplicative operators have the same multiplicative properties as of the continuum functions,~$[f][g] = [g][f] = [fg]$.

\subsection{Discrete equations: The SBP scheme}\label{sec:SBP_Scheme}

We now discretize the~EOMs,~\eqref{eq:hyp_LWEP_rescaled_FOR_null_Tilded}-\eqref{eq:hyp_LWEP_rescaled_FOR_null_Tilded_z-axis_scri}, and the energy norm \eqref{eq:Energy_norm}-\eqref{eq:Energy_Density_Rescaled_Variables}, to obtain the~SBP scheme. Most of these derivations remain independent of the choice of discretization, whether~FD or pseudo-spectral, and, thus, of the grid structure. We begin with the case of zero constraint damping,~$\tilde{\zeta} = 0$, and then mention how the scheme is modified when the constraint damping is switched on, for the case~$\tilde{\zeta} = 1$.

The discrete~EOMs in the bulk now become
\begin{align}\label{eq:Discrete_EOM_bulk}
\dot{\tilde{\Psi}} = &\ \frac{1}{2} \left( \left[ \frac{1}{\chi} \right] \tilde{\Psi}_+ + \tilde{\Psi}_- \right) \, , \non  \\
\dot{\tilde{\Psi}}_+ = &\ \left[ \frac{\chi}{2R'-1} \right] \Bigg( (\Upsilon^r)^{-1} \left[ \frac{D_r + \tilde{D}_r}{2} \right] \left[ \frac{1}{\chi} \right] \tilde{\Psi}_+ + (\Upsilon^r)^{-1} \left[ \frac{D_r - \tilde{D}_r}{2} \right] \tilde{\Psi}_- - \left[ \frac{\chi'}{\chi} \right] \tilde{\Psi}_- \non \\
& + \left[ \frac{R'}{R^2} \right] \bigg( (\Upsilon^\theta)^{-1} \, \tilde{D}_\theta \tilde{\Psi}_\theta + \left[ \frac{1}{\sin^2\theta} \right] (\Upsilon^\phi)^{-1} \, \tilde{D}_\phi \tilde{\Psi}_\phi \bigg) - \left[ R' F \right] \tilde{\Psi} + \tilde{\zeta} \left[ \frac{1}{\chi} \right] \bigg( (\Upsilon^r)^{-1} \, D_r \tilde{\Psi} \non \\
& - \left[ \frac{\chi'}{\chi} \right] \tilde{\Psi} - \left[ \frac{2R'-1}{2 \chi} \right] \tilde{\Psi}_+ + \frac{\tilde{\Psi}_-}{2} \bigg) \Bigg) \, , \non \\
\dot{\tilde{\Psi}}_- = &\ - (\Upsilon^r)^{-1} \left[ \frac{D_r + \tilde{D}_r}{2} \right] \tilde{\Psi}_- - (\Upsilon^r)^{-1} \left[ \frac{D_r - \tilde{D}_r}{2} \right] \left[ \frac{1}{\chi} \right] \tilde{\Psi}_+ + \left[ \frac{R'}{R^2} \right] \bigg(  (\Upsilon^\theta)^{-1} \, \tilde{D}_\theta \tilde{\Psi}_\theta \non \\
& + \left[ \frac{1}{\sin^2\theta} \right]  (\Upsilon^\phi)^{-1} \, \tilde{D}_\phi \tilde{\Psi}_\phi \bigg) + \left[ \frac{\chi'}{\chi^2} \right] \tilde{\Psi}_+ - \left[ R' F \right] \tilde{\Psi} - \tilde{\zeta} \left[ \frac{1}{\chi} \right] \bigg( (\Upsilon^r)^{-1} \, D_r \tilde{\Psi} - \left[ \frac{\chi'}{\chi} \right] \tilde{\Psi} \non \\
& - \left[ \frac{2R'-1}{2 \chi} \right] \tilde{\Psi}_+ + \frac{\tilde{\Psi}_-}{2} \bigg) \, , \non \\
\dot{\tilde{\Psi}}_\theta = &\ \frac{1}{2} \, (\Upsilon^\theta)^{-1} \, D_\theta \left( \left[ \frac{1}{\chi} \right] \tilde{\Psi}_+ + \tilde{\Psi}_- \right) + \tilde{\zeta} \left[ \frac{1}{\chi} \right] \bigg( (\Upsilon^\theta)^{-1} \, D_\theta \tilde{\Psi} - \tilde{\Psi}_\theta \bigg) \, , \non \\
\dot{\tilde{\Psi}}_\phi = &\ \frac{1}{2} \, (\Upsilon^\phi)^{-1} \, D_\phi \left( \left[ \frac{1}{\chi} \right] \tilde{\Psi}_+ + \tilde{\Psi}_- \right) + \tilde{\zeta} \left[ \frac{1}{\chi} \right] \bigg( (\Upsilon^\phi)^{-1} \, D_\phi \tilde{\Psi} - \tilde{\Psi}_\phi \bigg) \, ,
\end{align}
everywhere,
\begin{align}\label{eq:Discrete_EOM_bulk_z-axis}
\dot{\tilde{\Psi}} = &\ \frac{1}{2} \left( \left[ \frac{1}{\chi} \right] \tilde{\Psi}_+ + \tilde{\Psi}_- \right) \, , \non  \\
\dot{\tilde{\Psi}}_+ = &\ \left[ \frac{\chi}{2R'-1} \right] \Bigg( (\Upsilon^r)^{-1} \left[ \frac{D_r + \tilde{D}_r}{2} \right] \left[ \frac{1}{\chi} \right] \tilde{\Psi}_+ + (\Upsilon^r)^{-1} \left[ \frac{D_r - \tilde{D}_r}{2} \right] \tilde{\Psi}_- - \left[ \frac{\chi'}{\chi} \right] \tilde{\Psi}_- - \left[ R' F \right] \tilde{\Psi} \non \\
& + \left[ \frac{2 R'}{R^2} \right] (\Upsilon^\theta)^{-1} \, D_\theta \tilde{\Psi}_{\theta (m=0)} + \tilde{\zeta} \left[ \frac{1}{\chi} \right] \bigg( \frac{\tilde{\Psi}_-}{2} + (\Upsilon^r)^{-1} \, D_r \tilde{\Psi} - \left[ \frac{\chi'}{\chi} \right] \tilde{\Psi} - \left[ \frac{2R'-1}{2 \chi} \right] \tilde{\Psi}_+ \bigg) \Bigg)  \, , \non \\
\dot{\tilde{\Psi}}_- = &\ - (\Upsilon^r)^{-1} \left[ \frac{D_r + \tilde{D}_r}{2} \right] \tilde{\Psi}_- - (\Upsilon^r)^{-1} \left[ \frac{D_r - \tilde{D}_r}{2} \right] \left[ \frac{1}{\chi} \right] \tilde{\Psi}_+ + \left[ \frac{\chi'}{\chi^2} \right] \tilde{\Psi}_+ - \left[ R' F \right] \tilde{\Psi} \non \\
& + \left[ \frac{2 R'}{R^2} \right] (\Upsilon^\theta)^{-1} \, D_\theta \tilde{\Psi}_{\theta (m=0)} - \tilde{\zeta} \left[ \frac{1}{\chi} \right] \bigg( \frac{\tilde{\Psi}_-}{2} + (\Upsilon^r)^{-1} \, D_r \tilde{\Psi} - \left[ \frac{\chi'}{\chi} \right] \tilde{\Psi} - \left[ \frac{2R'-1}{2 \chi} \right] \tilde{\Psi}_+ \bigg) \, , \non \\
\dot{\tilde{\Psi}}_\theta = &\ \frac{1}{2} \, (\Upsilon^\theta)^{-1} \, D_\theta \left( \left[ \frac{1}{\chi} \right] \tilde{\Psi}_+ + \tilde{\Psi}_- \right) + \tilde{\zeta} \left[ \frac{1}{\chi} \right] \left( (\Upsilon^\theta)^{-1} \, D_\theta \tilde{\Psi} -\tilde{\Psi}_\theta \right) \, , \non \\
\dot{\tilde{\Psi}}_\phi = &\ 0 \, ,
\end{align}
on the~$z$-axis, and
\begin{align}\label{eq:Discrete_EOM_origin}
\dot{\tilde{\Psi}} = &\ \frac{\tilde{\Psi}_+ + \tilde{\Psi}_-}{2} \, , \non  \\
\dot{\tilde{\Psi}}_+ = &\ \frac{1}{2} \, (\Upsilon^r)^{-1} \, D_r \left( 3 \, (\tilde{\Psi}_+ - \tilde{\Psi}_-)_{(l=0)} + (\tilde{\Psi}_+ + \tilde{\Psi}_-) \right) - [F] \tilde{\Psi}_{(l=0)} + \tilde{\zeta} \, \bigg( (\Upsilon^r)^{-1} \, D_r \tilde{\Psi} - \frac{\tilde{\Psi}_+ - \tilde{\Psi}_-}{2} \bigg) \, , \non \\
\dot{\tilde{\Psi}}_- = &\ \frac{1}{2} \, (\Upsilon^r)^{-1} \, D_r \left( 3 \, (\tilde{\Psi}_+ - \tilde{\Psi}_-)_{(l=0)} - (\tilde{\Psi}_+ + \tilde{\Psi}_-) \right) - [F] \tilde{\Psi}_{(l=0)} - \tilde{\zeta} \, \bigg( (\Upsilon^r)^{-1} \, D_r \tilde{\Psi} - \frac{\tilde{\Psi}_+ - \tilde{\Psi}_-}{2} \bigg) \, , \non \\
\dot{\tilde{\Psi}}_\theta = &\ 0 \, , \quad
\dot{\tilde{\Psi}}_\phi = 0 \, ,
\end{align}
at the origin. The ones at~$\mathscr{I}^+$, for~$n=2$, become
\begin{align}\label{eq:Discrete_EOM_scri}
\dot{\tilde{\Psi}} = &\ \frac{\tilde{\Psi}_-}{2} \, , \quad
\dot{\tilde{\Psi}}_+ = - \frac{\tilde{\Psi}_-}{2} - \left[ \frac{R F}{2} \right] \tilde{\Psi} \, , \non \\
\dot{\tilde{\Psi}}_- = &\ - (\Upsilon^r)^{-1} \left[ \frac{D_r + \tilde{D}_r}{2} \right] \tilde{\Psi}_- - (\Upsilon^r)^{-1} \left[ \frac{D_r - \tilde{D}_r}{2} \right] \left[ \frac{1}{\chi} \right] \tilde{\Psi}_+ + 2 \, \tilde{\Psi}_+ + 2 \, \bigg( (\Upsilon^\theta)^{-1} \, \tilde{D}_\theta \tilde{\Psi}_\theta \non \\
& + \left[ \frac{1}{\sin^2\theta} \right] (\Upsilon^\phi)^{-1} \, \tilde{D}_\phi \tilde{\Psi}_\phi \bigg) - \left[ R' F \right] \tilde{\Psi} + 2 \, \tilde{\zeta} \, ( \tilde{\Psi} + \tilde{\Psi}_+ ) \, , \non \\
\dot{\tilde{\Psi}}_\theta = &\ \frac{1}{2} \, (\Upsilon^\theta)^{-1} \, D_\theta \tilde{\Psi}_- \, , \quad
\dot{\tilde{\Psi}}_\phi = \frac{1}{2} \, (\Upsilon^\phi)^{-1} \, D_\phi \tilde{\Psi}_- \, ,
\end{align}
everywhere, and
\begin{align}\label{eq:Discrete_EOM_z-axis_scri}
\dot{\tilde{\Psi}} = &\ \frac{\tilde{\Psi}_-}{2} \, , \quad
\dot{\tilde{\Psi}}_+ = - \frac{\tilde{\Psi}_-}{2} - \left[ \frac{R F}{2} \right] \tilde{\Psi} \, , \non \\
\dot{\tilde{\Psi}}_- = &\ - (\Upsilon^r)^{-1} \left[ \frac{D_r + \tilde{D}_r}{2} \right] \tilde{\Psi}_- - (\Upsilon^r)^{-1} \left[ \frac{D_r - \tilde{D}_r}{2} \right] \left[ \frac{1}{\chi} \right] \tilde{\Psi}_+ + 2 \, \tilde{\Psi}_+ + 4 \, (\Upsilon^\theta)^{-1} \, D_\theta \tilde{\Psi}_{\theta (m=0)} \non \\
& - \left[ R' F \right] \tilde{\Psi} + 2 \, \tilde{\zeta} \, ( \tilde{\Psi} + \tilde{\Psi}_+ ) \, , \non \\
\dot{\tilde{\Psi}}_\theta = &\ \frac{1}{2} \, (\Upsilon^\theta)^{-1} \, D_\theta \tilde{\Psi}_- \, , \quad
\dot{\tilde{\Psi}}_\phi = 0 \, ,
\end{align}
on the~$z$-axis. Here, we compute the~$m=0$ and~$l=0$ modes the same way we did in the continuum case, eqs.~\eqref{eq:m=0_mode} and~\eqref{eq:l=0_mode}. As stated above, extracting the~$m=0$ mode correctly from the relation
\begin{align}\label{eq:Discrete_m=0_mode}
(\Psi_{m=0})_{IJK} = \frac{1}{2 \pi} \sum_{K'=0}^{N_\phi - 1} (\Upsilon^\phi)_{KK'} \Psi_{IJK'} \, ,
\end{align}
requires that, for every grid point~$K$, located at~$\phi_K$, there exists a grid point~$K + N_\phi/2$ located at~$\phi_K + \pi$, for~$\phi_K < \pi$, or~$K - N_\phi/2$, located at~$\phi_K - \pi$, for~$K\Delta\phi \geq \pi$, so that all~$m>0$ modes are cancelled perfectly. This condition requires~$N_\phi$ to be even.

Similarly, the~$l=0$ mode, extracted correctly from the relation
\begin{align}\label{eq:Discrete_l=0_mode}
\hspace*{-2.0em} (\Psi_{l=0})_{IJK} = \frac{1}{4\pi} \sum_{\substack{J',J'' \\ =0}}^{N_\theta} \sum_{K'=0}^{N_\phi - 1} (\Upsilon^\theta)_{JJ'} \, (\Upsilon^\phi)_{KK'} \, [\sin\theta]_{J'J''} \, \Psi_{IJ''K'} \, ,
\end{align}
requires, additionally, that for every grid point~$J$, located at~$\theta_J$, there must be a grid point~$N_\theta - J$ located at~$\pi - \theta_J$, so that all the~$l>0$ modes are cancelled perfectly. The modes~$\Psi_{m=0}$ are, thereby, independent of~$K$, and~$\Psi_{l=0}$ are independent of both~$J$ and~$K$.

Inspired by~\eqref{eq:Energy_norm} and~\eqref{eq:Energy_Density_Rescaled_Variables}, we now define the discrete energy norm as
\begin{align}\label{eq:Discrete_Energy_Norm}
\hspace{-1.0em}\hat{E}(t) = &\ \frac{1}{2} \bigg( \tilde{\Psi}^T \Upsilon [F] \tilde{W} \tilde{\Psi} + \tilde{\Psi}_+^T \Upsilon \tilde{W}_+ \tilde{\Psi}_+ + \tilde{\Psi}_-^T \Upsilon \tilde{W}_- \tilde{\Psi}_- + \tilde{\Psi}_\theta^T \Upsilon \tilde{W}_\theta \tilde{\Psi}_\theta + \tilde{\Psi}_\phi^T \Upsilon \tilde{W}_\phi \tilde{\Psi}_\phi \bigg) \, ,
\end{align}
with
\begin{align}\label{eq:Discrete_Weights}
& \tilde{W} = \left[ \frac{R' \, R^2 \, \sin\theta}{\chi^2} \right] \, , \quad \tilde{W}_+ = \left[ \frac{(2R'-1) \, R^2 \, \sin\theta}{2\, \chi^4} \right] , \quad \tilde{W}_- = \left[ \frac{R^2 \, \sin\theta}{2 \, \chi^2} \right] , \non \\
& \tilde{W}_\theta = \left[ \frac{R' \, \sin\theta}{\chi^2} \right] , \quad \tilde{W}_\phi = \left[ \frac{R'}{\chi^2 \, \sin\theta} \right] .
\end{align}
We define all the weight matrices~$\tilde{W}$'s here as suggested in~\eqref{eq:Discrete_Coefficients}. Taking the time derivative and substituting the EOMs \eqref{eq:Discrete_EOM_bulk} with~$\tilde{\zeta} = 0$, and using the properties~\eqref{eq:Quadrature}-\eqref{eq:Discrete_Coefficients}, we obtain
\begin{align}
\p_t \hat{E}(t) = &\ \tilde{\Psi}_+^T \left[ \frac{1}{\chi} \right] \Upsilon^\theta \, \Upsilon^\phi \, \tilde{W}_- \left[ \frac{D_r + \tilde{D}_r}{2} \right] \left[ \frac{1}{\chi} \right] \tilde{\Psi}_+ - \tilde{\Psi}_-^T \, \Upsilon^\theta \, \Upsilon^\phi \, \tilde{W}_- \left[ \frac{D_r + \tilde{D}_r}{2} \right] \tilde{\Psi}_- \non \\
& + \tilde{\Psi}_+^T \left[ \frac{1}{\chi} \right] \Upsilon^\theta \, \Upsilon^\phi \, \tilde{W}_- \left[ \frac{D_r - \tilde{D}_r}{2} \right] \tilde{\Psi}_- - \tilde{\Psi}_-^T \, \Upsilon^\theta \, \Upsilon^\phi \, \tilde{W}_- \left[ \frac{D_r - \tilde{D}_r}{2} \right] \left[ \frac{1}{\chi} \right] \tilde{\Psi}_+ \non \\
& + \tilde{\Psi}_+^T \left[ \frac{1}{\chi} \right] \Upsilon^r \, \Upsilon^\phi \, \frac{\tilde{W}_\theta}{2} \, \tilde{D}_\theta \tilde{\Psi}_\theta + \tilde{\Psi}_\theta^T \, \Upsilon^r \, \Upsilon^\phi \, \frac{\tilde{W}_\theta}{2} D_\theta \left[ \frac{1}{\chi} \right] \tilde{\Psi}_+ + \tilde{\Psi}_-^T \, \Upsilon^r \, \Upsilon^\phi \, \frac{\tilde{W}_\theta}{2} \, \tilde{D}_\theta \tilde{\Psi}_\theta \non \\
& + \tilde{\Psi}_\theta^T \, \Upsilon^r \, \Upsilon^\phi \, \frac{\tilde{W}_\theta}{2} D_\theta \tilde{\Psi}_- + \tilde{\Psi}_+^T \left[ \frac{1}{\chi} \right] \Upsilon^r \, \Upsilon^\theta \, \frac{\tilde{W}_\phi}{2} \, \tilde{D}_\phi \tilde{\Psi}_\phi + \tilde{\Psi}_\phi^T \, \Upsilon^r \, \Upsilon^\phi \, \frac{\tilde{W}_\phi}{2} D_\phi \left[ \frac{1}{\chi} \right] \tilde{\Psi}_+ \non \\
& + \tilde{\Psi}_-^T \, \Upsilon^r \, \Upsilon^\theta \, \frac{\tilde{W}_\phi}{2} \, \tilde{D}_\phi \tilde{\Psi}_\phi + \tilde{\Psi}_\phi^T \, \Upsilon^r \, \Upsilon^\phi \, \frac{\tilde{W}_\phi}{2} D_\phi \tilde{\Psi}_- \, .
\end{align}
The~SBP scheme is now derived by requiring that this energy change equals a flux at the outer boundary that approaches~\eqref{eq:E-dot_rescaled} in the continuum limit. Its discrete analogue can be expressed as
\begin{align}\label{eq:Boundary_Matrix_Definition}
\p_t \hat{E}(t) = \tilde{\Psi}_+^T \left[ \frac{1}{\chi} \right] \Upsilon^\theta \, \Upsilon^\phi \, B \left[ \frac{1}{\chi} \right] \tilde{\Psi}_+ - \tilde{\Psi}_-^T \, \Upsilon^\theta \, \Upsilon^\phi \, B \, \tilde{\Psi}_- \, ,
\end{align}
where~$B$ is called the `boundary matrix', which is nontrivial only at the boundary points, i.e.~$B_{II'} = 0$ for both~$I,I' < N_r$. The quadratic from~\eqref{eq:Boundary_Matrix_Definition} also suggests that one can take~$B$ to be symmetric
\begin{align}
B_{IJKI'J'K'} = B_{I'JKIJ'K'} = B_{IJ'KI'JKK'} = B_{IJK'I'J'K} \, ,
\end{align}
without losing generality. This condition gives us the following relations
\begin{align}\label{eq:SBP_Scheme}
& \tilde{W}_- \left[ \frac{D_r + \tilde{D}_r}{2} \right] + \left( \tilde{W}_- \left[ \frac{D_r + \tilde{D}_r}{2} \right] \right)^T = 2\, B \, , \non \\
& \tilde{W}_- \left[ \frac{D_r - \tilde{D}_r}{2} \right] - \left( \tilde{W}_- \left[ \frac{D_r - \tilde{D}_r}{2} \right] \right)^T = 0 \, , \non \\
& \tilde{W}_\theta \, \tilde{D}_\theta + (\tilde{W}_\theta \, D_\theta)^T = 0 \, , \quad \tilde{W}_\phi \, \tilde{D}_\phi + (\tilde{W}_\phi \, D_\phi)^T \, .
\end{align}
Here, we used the fact that~$\Upsilon^\theta$ and~$\Upsilon^\phi$ are scalars along the~$\hat{r}$-direction, which allows them to commute with all radial operators. Similarly,~$\Upsilon^r$ and~$\Upsilon^\phi$ commute with all operators along the~$\hat{\theta}$-direction, while~$\Upsilon^r$ and~$\Upsilon^\theta$ commute with those along the~$\hat{\phi}$-direction. For the same reason, the quadratures~$\Upsilon^r$,~$\Upsilon^\theta$, and~$\Upsilon^\phi$ also commute with each other. The system~\eqref{eq:SBP_Scheme} constitutes what we refer to as our final \textit{\textbf{summation-by-parts~(SBP) scheme}}.

We next solve these equations for various operators. Since the~$\tilde{W}$'s do not have any~$\phi$-dependence,~$\tilde{W}_\phi$ commutes with~$D_\phi$ and~$\tilde{D}_\phi$, and the last equation in~\eqref{eq:SBP_Scheme} simply gives
\begin{align}
\tilde{D}_\phi = - D_\phi^T \, .
\end{align}
This equation gives~$\tilde{D}_\phi = D_\phi$ whenever~$(D_\phi)^T = - D_\phi$. This result is consistent with the continuum limit~\eqref{eq:Tilded_Operators}. We shall demonstrate this construction in the next subsection.

The third equation gives
\begin{align}\label{eq:Dtilde_theta_abstract}
\tilde{D}_\theta = - \tilde{W}_\theta^{-1} \, D_\theta^T \, \tilde{W}_\theta \, .
\end{align}
Since all radial operators~$[f(r)]$ are scalars along~$\hat{\theta}$, they commute with~$D_\theta$ and~$\tilde{D}_\theta$, giving the simple relation
\begin{align}\label{eq:Dtilde_theta}
\tilde{D}_\theta = - \left[ \frac{1}{\sin\theta} \right] D_\theta^T \bigg[ \sin\theta \bigg] \, .
\end{align}
As we observe in~\eqref{eq:hyp_LWEP_rescaled_FOR_null_Tilded_z-axis},~\eqref{eq:hyp_LWEP_rescaled_FOR_null_Tilded_z-axis_scri},~\eqref{eq:Discrete_EOM_bulk_z-axis} and~\eqref{eq:Discrete_EOM_z-axis_scri}, both~$\tilde{\p}_\theta$ and~$\tilde{D}_\theta$ do not appear in the equations on the~$z$-axis, we do not need to worry about their singular behavior there. Again, if we choose~$D_\theta$ such that~$(D_\theta)^T = - D_\theta$, we obtain the relation
\begin{align}\label{eq:Dtilde_theta_applied}
\tilde{D}_\theta = \left[ \frac{1}{\sin\theta} \right] D_\theta \bigg[ \sin\theta \bigg] \, ,
\end{align}
which is consistent with~\eqref{eq:Tilded_Operators}, as we shall observe shortly.

The first two equations give
\begin{align}\label{eq:SBP_Scheme_1}
\tilde{D}_r = - \tilde{W}_-^{-1} \, D_r^T \, \tilde{W}_- + 2 \, \tilde{W}_-^{-1} \, B \, ,
\end{align}
with two unknowns,~$\tilde{D}_r$ and~$B$. Since all these operators are radial, any angular dependence in~$\tilde{W}_-$ will commute with them and cancel out, leaving
\begin{align}
\tilde{D}_r = - \left[ \frac{\chi^2}{R^2} \right] D_r^T \left[ \frac{R^2}{\chi^2} \right] + 2\, \tilde{W}_-^{-1} \, B \, .
\end{align}
Any symmetric~$B$, coming from the first equation in \eqref{eq:SBP_Scheme}, gives a~$\tilde{D}_r$ that does not necessarily converge to~$\tilde{\p}_r$, defined in~\eqref{eq:Tilded_Operators}, in the continuum limit, unless we decrease the accuracy of~$D_r$ at the outer boundary. To address this issue, we define~$\tilde{D}_r$ analogous to the continuum definition
\begin{align}\label{eq:Dtilde_r}
\tilde{D}_r = \left[ \frac{\chi^2}{R^2} \right] D_r \left[ \frac{R^2}{\chi^2} \right] \, ,
\end{align}
and then solve~\eqref{eq:SBP_Scheme_1} to get
\begin{align}\label{eq:Boundary_Matrix}
B = \frac{1}{2} \, \tilde{W}_- \left[ \frac{\chi^2}{R^2} \right] (D_r + D_r^T) \left[ \frac{R^2}{\chi^2} \right] \, .
\end{align}
The price we pay is that~$B$ is no longer symmetric for a general~$D_r$. Consequently, the right sides of the first two equations in~\eqref{eq:SBP_Scheme} are replaced by~$B^T+B$ and~$B^T-B$, respectively. However, both~$B^T+B$ and~$B^T-B$ can be made to converge to the continuum limit at the desired accuracy. We shall see it shortly.

Equation~\eqref{eq:Boundary_Matrix} gives~$B_{II'} = 0$ for~$I,I' < N_r$ whenever~$D_r$ is antisymmetric in the bulk, and gives nontrivial~$B_{IN_r}$ and $B_{N_r I}$, at least for the last few indices~$I$ in~$[0, N_r]$, depending on the definition of~$D_r$ at the outer boundary. Such a construction is possible for an~FD scheme, as demonstrated in the next subsection.

Lastly, all the results above remain unaltered even with constraint damping, and so, constraint addition does not affect the~SBP scheme.

\subsection{Discrete Operators and Stencils}\label{sec:Disc_Ops}

We choose second-order accurate finite-difference~(FD) operators for numerical implementation, defined as
\begin{align}\label{eq:Dr_Bulk}
& \left( D_r f \right)_{I,J,K} = \frac{f_{I+1,J,K} - f_{I-1,J,K}}{2} \, , \quad \textrm{for } I = 1, \ldots, N_r-1 \quad \textrm{and } \forall \;\; J \textrm{ and } K \, , \non \\
& \left( D_\theta f \right)_{I,J,K} = \frac{f_{I,J+1,K} - f_{I,J-1,K}}{2} \, , \quad \textrm{for } J = 1, \ldots, N_\theta-1 \quad \textrm{and } \forall \;\; I \textrm{ and } K \, , \non \\
& \left( D_\phi f \right)_{I,J,K} = \frac{f_{I,J,K+1} - f_{I,J,K-1}}{2} \, , \quad \textrm{for } K = 0, \ldots, N_\phi-1 \quad \textrm{and } \forall \;\; I \textrm{ and } J \, ,
\end{align}
To define these operators at the endpoints, we extend the computational grid virtually across the domain by introducing ghost points in the regions~$r<0$,~$\theta < 0$,~$\theta > \pi$,~$\phi < 0$, and~$\phi \geq 2\pi$, with the same grid width~$\Delta r$, $\Delta \theta$ and $\Delta \phi$, and populate them using the parity conditions~\eqref{eq:Parity}, \eqref{eq:Parity_FOR_R_theta}, \eqref{eq:Parity_phi}, \eqref{eq:Parity_FOR_Null_R} and \eqref{eq:Parity_Chi}. To ensure that~$(\Upsilon^r)^{-1} D_r$,~$(\Upsilon^\theta)^{-1} D_\theta$ and~$(\Upsilon^\phi)^{-1} D_\phi$ remain discrete approximations for~$\p_r$,~$\p_\theta$ and~$\p_\phi$ everywhere, respectively, we define these operators at the endpoints as
\begin{align}\label{eq:FD_Endpoints}
& \left( D_r f \right)_{0,J,K} = \frac{f_{1,J,K} - f_{-1,J,K}}{4} \, , \quad \forall \, J,K \, , \non \\
& \left( D_\theta f \right)_{I,0,K} = \frac{f_{I,1,K} - f_{I,-1,K}}{4} \, , \quad \forall \, I,K \, , \quad \textrm{and} \non \\
& \left( D_\theta f \right)_{I,N_\theta,K} = \frac{f_{I,N_\theta + 1,K} - f_{I,N_\theta - 1,K}}{4} \, , \quad \forall \, I,K \, .
\end{align}

We observe that~$D_\phi$, as defined in~\eqref{eq:Dr_Bulk}, is antisymmetric throughout the domain, giving~$\tilde{D}_\phi = D_\phi$. Similarly, the antisymmetry of~$D_\theta$, as defined in~\eqref{eq:Dr_Bulk} and~\eqref{eq:FD_Endpoints}, leads to~\eqref{eq:Dtilde_theta_applied}. Interestingly, the operator~$D_r$, defined in~\eqref{eq:Dr_Bulk} and~\eqref{eq:FD_Endpoints}, is also antisymmetric in the bulk,~$0 < I < N_r$, resulting in~$B_{II'} = 0$ for~$0 < I,I' < N_r$. This is consistent with our requirement for~$B$ to be a boundary matrix, cf.~\eqref{eq:Boundary_Matrix_Definition}. Since~$\tilde{D}_r$ does not appear in the equations at the origin, we do not concern ourselves with its definition and properties at~$I=0$. The only remaining task is to define~$D_r$ at~$I=N_r$. Since this point corresponds to the actual boundary of the domain, we need to define it here using upwinding, as performed in~\cite{GauVanHil21}. This leads to two kinds of~SBP schemes: \textit{SBP-TEM} (for \textit{truncation-error-matching}) and \textit{SBP-stable}, depending on the stencil at~$I=N_r$.

\subsubsection{SBP-TEM}\label{sec:SBP_TEM}

Here, we define~$D_r$ at~$I = N_r$ by imposing the~TEM property~\cite{Pre02}, resulting in
\begin{align}\label{eq:Dr_TEM}
\left( D_r f \right)_{N_r, J, K} = &\ \frac{1}{4} \, \bigl( f_{N_r-4, J, K} - 5 \, f_{N_r-3, J, K} + 10 \, f_{N_r-2, J, K} - 11 \, f_{N_r-1, J, K} + 5 \, f_{N_r, J, K} \bigl) \, ,
\end{align}
for all~$J$ and~$K$. Ref.~\cite{GauVanHil21, GauRedKum26} for more details. The resulting boundary matrix becomes
\begin{align}
B = \left( \begin{array}{ccccccc}
\ldots & . & . & . & . & . & . \\
\ldots & 0 & 0 & 0 & 0 & 0 & 0 \\
\ldots & 0 & 0 & 0 & 0 & 0 & \frac{W_{N_r N_r}}{8} \\
\ldots & 0 & 0 & 0 & 0 & 0 & -\frac{5 W_{N_r N_r}}{8} \\
\ldots & 0 & 0 & 0 & 0 & 0 & \frac{5 W_{N_r N_r}}{4} \\
\ldots & 0 & 0 & 0 & 0 & 0 & -\frac{9 W_{N_r N_r}}{8} \\
\ldots & 0 & \frac{W_{(N_r-4) (N_r-4)}}{8} & -\frac{5 W_{(N_r-3) (N_r-3)}}{8} & \frac{5 W_{(N_r-2) (N_r-2)}}{4} & -\frac{9 W_{(N_r-1) (N_r-1)}}{8} & \frac{5 W_{N_r N_r}}{4}
\end{array} \right) \, ,
\end{align}
which is neither diagonal nor positive definite in its symmetric part. Consequently,~$\dot{\hat{E}}(t)$ becomes positive whenever the state corresponding to the negative eigenvalue dominates over the others. However, this state corresponds to highly noisy data near~$\mathscr{I}^+$, which can be mitigated by introducing dissipation. Nonetheless, the resulting flux at~$\mathscr{I}^+$ approaches the continuum limit~\eqref{eq:Stokes_diff_form_rescaled} with second-order accuracy. Ref.~\cite{GauVanHil21} for more details.

\subsubsection{SBP-Stable}\label{sec:SBP_Stable}

Here, we define~$D_r$ at the boundary such that it makes $B$ diagonal. For second-order accuracy in the bulk, the only choice that ensures this structure is
\begin{align}\label{eq:Dr_Stable}
\hspace*{-1.0em} (D_r f)_{N_r,J,K} = \frac{f_{N_r,J,K} - f_{N_r-1,J,K}}{2} \, , 
\quad \forall \, J \textrm{ and } K \, ,
\end{align}
resulting in the following boundary matrix
\begin{align}
B = {\rm diag}(0, \ldots, 0, (\tilde{W}_-)_{N_r N_r}/2) \, ,
\end{align}
Although this approach reduces accuracy at the outer boundary, it ensures that~$\dot{\hat{E}}(t)$ remains negative semi-definite for all times~$t$. Ref.~\cite{GauVanHil21} for a more details.

\subsection{Constraint Damping}\label{sec:Constraints}

The discrete~FOR constraints are now given by
\begin{align}
\hat{C}_r \equiv &\ \bigg( \left[ \frac{1}{\chi} \right] (\Upsilon^r)^{-1} \, D_r \tilde{\Psi} - \left[ \frac{\chi'}{\chi^2} \right] \tilde{\Psi} - \left[ \frac{2R'-1}{2 \chi^2} \right] \tilde{\Psi}_+ + \left[ \frac{1}{2 \chi} \right] \tilde{\Psi}_- \bigg) \, , \non \\
\hat{C}_{\theta} \equiv &\ \Upsilon_{\theta}^{-1}D_{\theta}\tilde{\Psi} - \tilde{\Psi}_{\theta} \, , \quad
\hat{C}_{\phi} \equiv \Upsilon_{\phi}^{-1}D_{\phi}\tilde{\Psi} - \tilde{\Psi}_{\phi} \, ,
\end{align}
and the discrete~EOMs lead to the solution~$\dot{\hat{C}}_i = - \zeta \, \hat{C}_i$, with the subscript~`$i$' denoting the~$r$,~$\theta$,~$\phi$ components, giving the overall solution~$\hat{C}_i (t) = \hat{C}_i (0) \, e^{-\zeta \, t}$, as in the continuum case. Therefore,~$\tilde{\zeta} = 0$ again gives the solution~$\hat{C}(t) = \hat{C}(0)$, for all~$t$.

The discrete constraints differ from the continuum ones at the second-order accuracy,
\begin{align}
& \hat{C}_r (t) = C_r (t) + O((\Delta r)^2) \, , \quad
 \hat{C}_\theta (t) = C_\theta (t) + O((\Delta \theta)^2) \, , \quad
 \hat{C}_\phi (t) = C_\phi (t) + O((\Delta \phi)^2) \, ,
\end{align}
and at the first-order accuracy in the~SBP-Stable scheme for~$\hat{C}_r$ at~$I = N_r$.

\subsection{Dissipation}\label{sec:Dissipation}

As concluded in~\cite{GauRedKum26}, we need to introduce dissipation only in the equation for~$\psi_T$, acting on~$\psi_T$, modifying the overall system to
\begin{align}\label{eq:LWEP_FOR_Constraint_damping_dissipation}
\p_T \psi = &\ \psi_T \, , \non \\
\p_T \psi_T = &\ - F \psi + \frac{1}{R^2} \p_R \left( R^2 \, \psi_R \right)
 + \frac{1}{R^2 \sin \theta} \p_\theta \left( \sin\theta \: \psi_\theta \right) + \frac{1}{R^2 \sin^2 \theta} \p_\phi \psi_\phi + Q  \, \psi_T \, , \non \\
\p_T \psi_R = &\ \p_R \psi_T + \zeta \, (\p_R \psi - \psi_R) \, , \quad
\p_T \psi_\theta = \p_\theta \psi_T + \zeta \, (\p_\theta \psi -  \psi_\theta) \, , \non \\
\p_T \psi_\phi = &\ \p_\phi \psi_T + \zeta \, (\p_\phi \psi -  \psi_\phi) \, .
\end{align}
This indicates that the~$4$-vector composed of the $(\psi_T, \psi_R, \psi_\theta, \psi_\phi)$-equations, should be added with the~$4$-vector~$C_{\mu'} \equiv (Q \, \psi_T, \gamma \, C_{i'})$, giving it an overall geometric structure.

The discrete system~\eqref{eq:Discrete_EOM_bulk}-~\eqref{eq:Discrete_EOM_z-axis_scri} is now modified as
\begin{align}\label{eq:Discrete_EOM_with_Diss_bulk}
\dot{\tilde{\Psi}}_+ = &\ \ldots + \left[ \frac{\chi}{2R'-1} \right] \tilde{Q} \left( \left[ \frac{1}{\chi} \right] \tilde{\Psi}_+ + \tilde{\Psi}_-  \right) \, , \quad
\dot{\tilde{\Psi}}_- = \ldots + \tilde{Q} \left( \left[ \frac{1}{\chi} \right] \tilde{\Psi}_+ + \tilde{\Psi}_- \right) \, ,
\end{align}
keeping the remaining equations unaltered. The ellipses `$\ldots$' here represent the right sides of the corresponding equations in the systems \eqref{eq:Discrete_EOM_bulk}-~\eqref{eq:Discrete_EOM_z-axis_scri}, and~$\tilde{Q}$ denotes a regularised form of~$Q$ on hyperboloidal slices, and the above equations suggest that it should remain~$O(1)$ everywhere.

As suggested in~\cite{GauRedKum26}, the fourth-order Kreiss-Oliger dissipation operator~(KODO)~\cite{GusKreOli95, KreOli73, KreLor89} can be defined for the~SBP-TEM scheme, to include the boundary points, as
\begin{align}\label{eq:Diss_TEM_L2_Norm}
\hspace{-1.0em} Q^{(4)}_1 \equiv \left\{ \begin{array}{l}
- a \, (\Upsilon^r)^3 \, \left( (\Upsilon^r)^{-1} D_+ \right)^2 \, \left( (\Upsilon^r)^{-1} D_- \right)^2 \, , \\
\vspace{0.5em} \textrm{for } I = 0, \ldots, N_r - 2 \, , \\
- a \, (\Upsilon^r)^3 \, (\Upsilon^r)^{-1} D_+ \, \left( (\Upsilon^r)^{-1} D_- \right)^3 \, , \\
\vspace{0.5em} \textrm{for } I = N_r - 1 \, , \textrm{and, } \\
- a \, (\Upsilon^r)^3 \, \left( (\Upsilon^r)^{-1} D_- \right)^4 \, , \textrm{ for } I = N_r \, ,
\end{array} \right.
\end{align}
where
\begin{align}\label{eq:Dplus_minus}
(D_\pm f)_I \equiv \pm (f_{I \pm 1} - f_I) \, ,
\end{align}
giving the following matrix representation near the outer boundary
\begin{align}
\hspace{-1.0em} Q^{(4)}_1 = \frac{- a}{(\Delta r)}
\left( \begin{array}{cccccccc}
. & . & . & . & . & . & . & . \\
. & 6 & -4 & 1 & 0 & 0 & 0 & 0 \\
. & -4 & 6 & -4 & 1 & 0 & 0 & 0 \\
. & 1 & -4 & 6 & -4 & 1 & 0 & 0 \\
. & 0 & 1 & -4 & 6 & -4 & 1 & 0 \\
. & 0 & 0 & 1 & -4 & 6 & -4 & 1 \\
. & 0 & 0 & 1 & -4 & 6 & -4 & 1 \\
. & 0 & 0 & 1 & -4 & 6 & -4 & 1 \\
\end{array} \right) .
\end{align}

With the SBP-stable scheme, we impose the dissipative property~(DP)~\cite{CalLehReu03, GauVanHil21}
\begin{align}
& (\Psi ,Q^{(4)} \, \Psi) \equiv - a \, \Psi^T \, \Upsilon^r \, Q^{(4)} \, \Psi = - a \left( (\Upsilon^r)^2 \left( (\Upsilon^r)^{-1} D_-)^2 \Psi \right) \right)^T \left( (\Upsilon^r)^2 \left( (\Upsilon^r)^{-1} D_-)^2 \Psi \right) \right) ,
\end{align}
which gives
\begin{align}\label{eq:Diss_stable_L2_Norm}
Q^{(4)}_2 = - a \, (\Upsilon^r)^{-1} \, D_-^T \, (\Upsilon^r)^{-1} \, D_-^T \, (\Upsilon^r)^2 \, D_- \, (\Upsilon^r)^{-1} \, D_- \, ,
\end{align}
and the following matrix representation near the outer boundary
\begin{align}
Q^{(4)}_2 = \frac{- a}{(\Delta r)}
\left( \begin{array}{cccccccc}
. & . & . & . & . & . & . & . \\
. & 6 & -4 & 1 & 0 & 0 & 0 & 0 \\
. & -4 & 6 & -4 & 1 & 0 & 0 & 0 \\
. & 1 & -4 & 6 & -4 & 1 & 0 & 0 \\
. & 0 & 1 & -4 & 6 & -4 & 1 & 0 \\
. & 0 & 0 & 1 & -4 & \frac{81}{16} & -\frac{17}{8} & \frac{1}{16} \\
. & 0 & 0 & 0 & 1 & -\frac{17}{8} & \frac{5}{4} & -\frac{1}{8} \\
. & 0 & 0 & 0 & 0 & \frac{1}{8} & -\frac{1}{4} & \frac{1}{8} \\
\end{array} \right) .
\end{align}

In our coordinate system~\cite{GauRedKum26}, we define dissipation as a discrete approximation of
\begin{align}
\tilde{Q}^{(4)} \equiv \left( \tilde{\p}_r \, \p_r + \frac{\chi^2}{R^2} \left( \tilde{\p}_\theta \, \p_\theta + \frac{1}{\sin^2\theta} \, \p_\phi^2 \right) \right)^2 \, ,
\end{align}
which can be defined for the~SBP-TEM and the~SBP-stable schemes as
\begin{align}
\tilde{Q}^{(4)}_{\rm TEM} \equiv &\ - a \, L^2 \, , \quad \tilde{Q}^{(4)}_{\rm stable} \equiv - a \, W_-^{-1} \, \Upsilon^{-1} \, L^T \, \Upsilon \, W_- \, L \, ,
\end{align}
respectively, where
\begin{align}
L \equiv &\ \bigg( (\Upsilon^r)^{-1} \, \tilde{D}_r \, (\Upsilon^r)^{-1} \, D_r + \left[ \frac{\chi^2}{R^2} \right] \bigg( (\Upsilon^\theta)^{-1} \, \tilde{D}_\theta \, (\Upsilon^\theta)^{-1} \, D_\theta + \left[ \frac{1}{\sin^2\theta} \right] ( (\Upsilon^\phi)^{-1} \, D_\phi)^2 \bigg) \bigg) \non \\
= &\ W_-^{-1} \bigg( (\Upsilon^r)^{-1} \, D_r \, W_- \, (\Upsilon^r)^{-1} \, D_r + \left[ \frac{\chi^2}{R^2} \right] \bigg( (\Upsilon^\theta)^{-1} \, D_\theta \, W_- \, (\Upsilon^\theta)^{-1} \, D_\theta \non \\
& + \left[ \frac{1}{\sin^2\theta} \right] (\Upsilon^\phi)^{-1} \, D_\phi \, W_- \, (\Upsilon^\phi)^{-1} \, D_\phi \bigg) \bigg) \, .
\end{align}
The definition of~$\tilde{Q}$ on the~$z$-axis and at the origin can be extended as in~\eqref{eq:LWEP_Sph_Coord_z-axis_m=0} and~\eqref{eq:LWEP_Sph_Coord_origin_l=0} for the~$\Box$ operator, respectively. Ref.~\cite{GauRedKum26} for more details.

\subsection{Time Integration}\label{sec:Time_Int}

We use the standard fourth-order Runge-Kutta (RK4) method for time integration. The Courant-Friedrichs-Lewy (CFL) condition in our choice of coordinates and grid is given by
\begin{align}
\Delta t = &\ \text{CFL} \times \min \left( \Delta r_I, \Delta r_I \, \Delta\theta_J, \Delta r_I \, \sin(\Delta \theta_J) \, \Delta\phi_K \right) \non \\
= &\ \text{CFL} \times \min \left( \frac{1}{N_r}, \frac{1}{N_r \, N_\theta}, \frac{1}{N_r \, N_\phi} \sin \left( \frac{1}{N_\theta} \right) \right) \, .
\end{align}

\begin{figure}[t]
\centering
\includegraphics[scale=1.0]{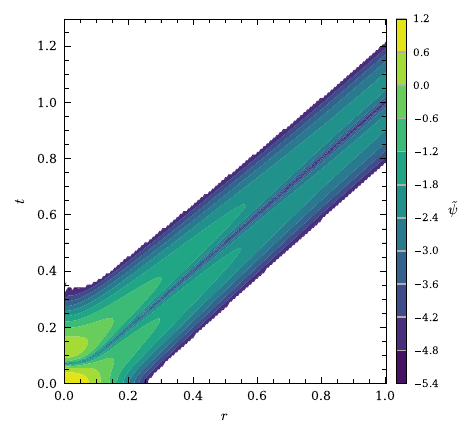}
\caption{Contour plot demonstrating the propagation of a narrow pulse of the scalar field~$\tilde{\psi}$ satisfying the~LWE towards~$\mathscr{I}^+$, located at $r = 1$. This solution is computed using the SBP-TEM scheme, and the corresponding SBP-Stable evolution exhibits qualitatively identical behavior. For clarity, the plot has been truncated at low amplitudes.}
\label{fig:Contour_TEM}
\end{figure}

\begin{figure}[t]
\centering
\includegraphics[width=0.7\linewidth]{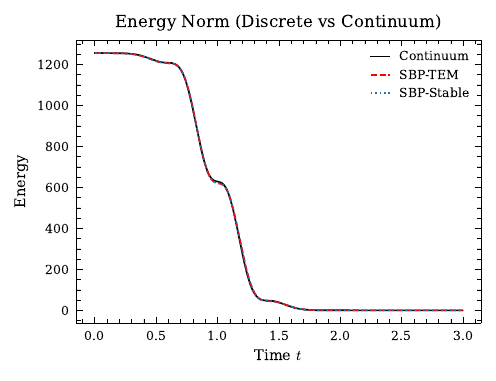}
\caption{Comparison of the continuum and discrete energies over time. The discrete energies are computed from the numerical data, obtained using both the~SBP-TEM and~SBP-Stable schemes. Both energy norms are non-increasing and closely align with the continuum energy calculated from a closed-form solution~\eqref{eq:psi_00_10_22_analytical}.}
\label{fig:energy_norm_disc_cont}
\end{figure}

\begin{figure*}[t]
\centering
\includegraphics[width=1.0\linewidth]{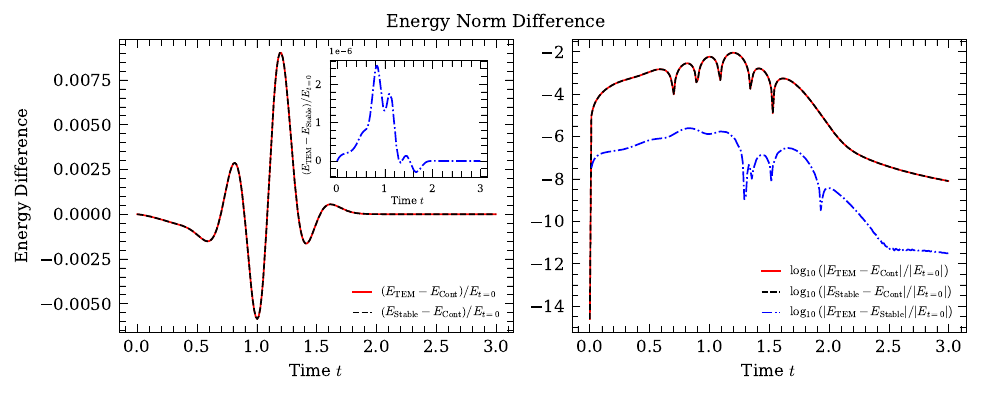}
\caption{Differences between the energy norms shown in Fig.~\ref{fig:energy_norm_disc_cont} over time. The differences between the discrete and continuum norms remain the same in both TEM and stable schemes upto~$4$ orders of magnitude.}
\label{fig:energy_norm_disc_cont_diff}
\end{figure*}

\begin{figure*}[t]
\centering
\includegraphics[width=1.0\linewidth]{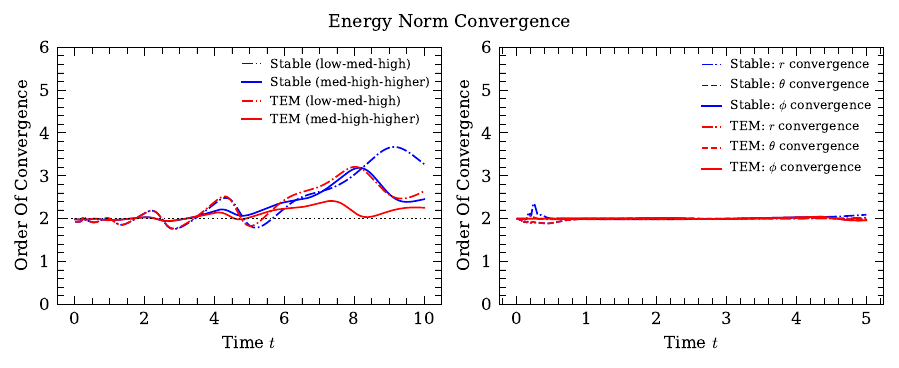}
\caption{Convergence order in the energy norm over time for the case~$F = 0$, and with the~ID specified in~\eqref{eq:psi_ID_00_20_22}. \emph{Left panel}: Convergence order both in~SBP-TEM and~SBP-Stable discretizations using four progressively refined uniform resolutions, approaching~$2$ at higher base resolutions. \emph{Right panel}: Convergence order along the three coordinate directions separately, which remains close to~$2$ for long times.}
\label{fig:norm_conv_TEM_Stable}
\end{figure*}

\section{Numerical Implementation and Results}\label{sec:numerics}

This section demonstrates the stability, energy conservation, convergence, and physical behavior of solutions propagating toward $\mathscr{I}^+$ through a series of numerical experiments, for the specific cases of~$F = 0$,~$F = 1/\chi^2$, and~$F = m^2$, utilizing both SBP-TEM and SBP-Stable discretization schemes. Some notable results from this study include:
\begin{enumerate}
\item Each simulation maintains stable evolution across all considered resolutions, even with a Courant factor as high as~CFL $= 2.6785$. This is the maximum value we found empirically.
\item All numerical evolutions remain stable even without dissipation,~$a = 0$, despite the presence of numerical noise.
\item Our system requires only minimal amounts of artificial dissipation to reduce high-frequency numerical noise:~$a = 0.002$ in the~SBP-TEM discretization, and~$a = 0.008$ for the SBP-Stable one.
\item As expected, incorporating dissipative terms into the equations for~$\tilde{\Psi}_\theta$ and~$\tilde{\Psi}_\phi$, as well as adding the dissipation operator acting on the linear combination~$\tilde{\Psi}_R \equiv \left( \left[ \frac{1}{\chi} \right] \tilde{\Psi}_+ - \tilde{\Psi}_- \right)$  to the equations for~$\tilde{\Psi}_\pm$ equations, does not influence the numerical noise in these directions. All these dissipation operators are defined in~\cite{GauRedKum26}.
\end{enumerate}

All the results shown below are generated from the~ID
\begin{align}\label{eq:psi_ID_00_20_22}
& \psi(0, R, \theta, \phi) = a \, e^{-\sigma^2 R^2} (1 + R^2 (\cos^2\theta - \sin^2\theta \, \sin(2\phi))) \, , \quad
 \psi_T (0, R, \theta, \phi) = 0 \, ,
\end{align}
with~$a = \sigma = 1$, unless stated otherwise. The~ID for all the variables~$\tilde{\Psi}$,~$\tilde{\Psi}_\pm$,~$\tilde{\Psi}_\theta$ and~$\tilde{\Psi}_\phi$ can be calculated from the definitions~\eqref{eq:FOR_TR_comps},~\eqref{eq:FOR_null_comps} and~\eqref{eq:Rescaled_FOR_null_comps}.

\begin{figure*}[t]
\centering
\includegraphics[width=\textwidth]{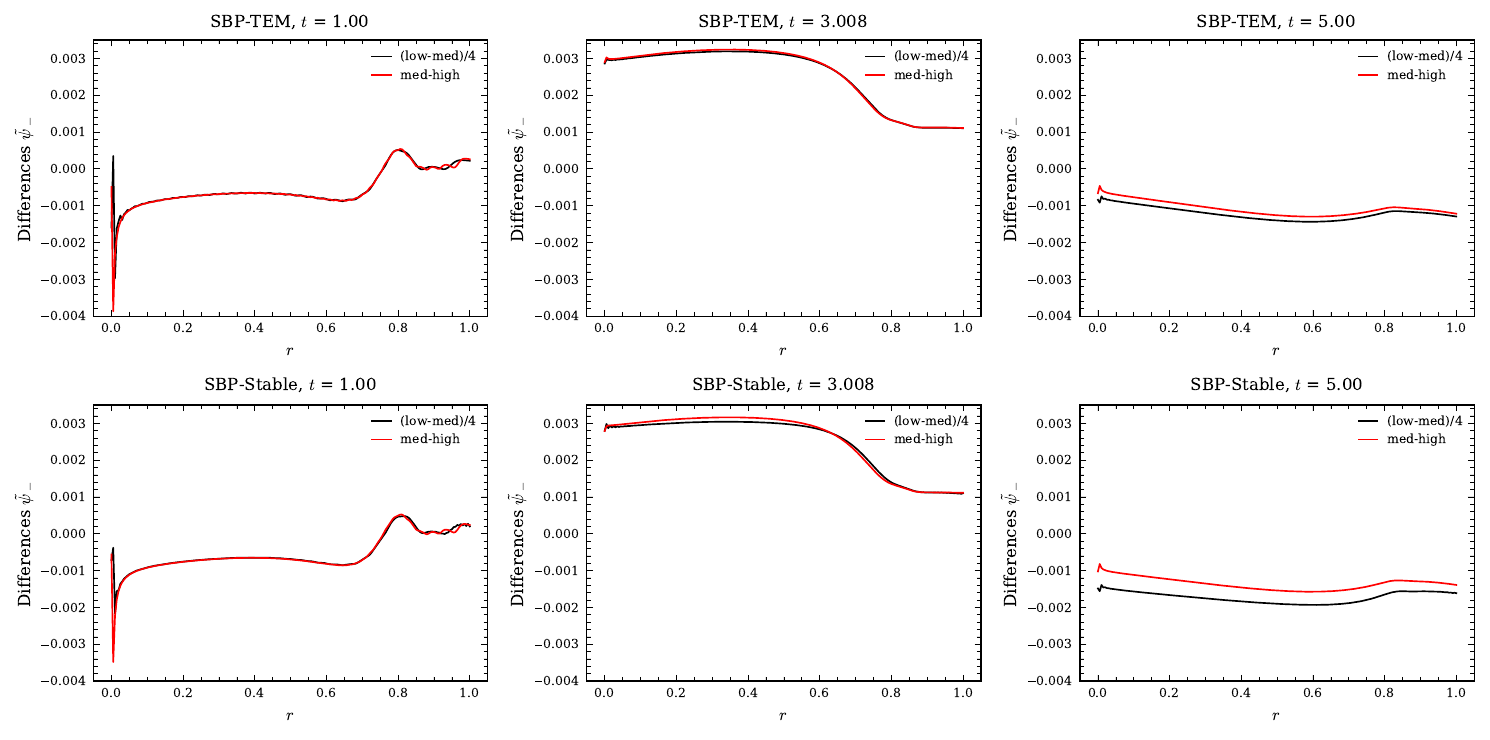}
\caption{Pointwise convergence along the radial direction~$\tilde{\psi}_-$, integrated over~$\theta$ and~$\phi$, showing a second-order convergence in both~SBP-TEM and~SBP-Stable schemes at various times. Spikes at the origin originate from the coordinate singularity.}
\label{fig:r_convergence_psim}
\end{figure*}

\subsection{Linear Wave Equation, $F=0$}\label{sec:F=0}

We begin by visualizing the evolution of a narrow pulse in the~ID, defined by eq.~\eqref{eq:psi_ID_00_20_22}with parameters~$a = 1$ and~$\sigma = 10$, evolves over time. Fig.~\ref{fig:Contour_TEM} shows that this pulse divides into outgoing and incoming components, which eventually arrive at~$\mathscr{I}^+$ in two distinct bursts. This behavior arises from the chosen time symmetry,~$\psi_T(T = 0) = 0$, in the~ID, ensuring that both outgoing and incoming components are treated equally. The outgoing velocity of these pulses equals unity, which aligns with our construction; cf. Sec.~\eqref{sec:Hyperboloidal_slices} for more details.

Figs.~\ref{fig:energy_norm_disc_cont} and~\ref{fig:energy_norm_disc_cont_diff} compare the total energy over time obtained from a numerical solution with the analytical solution for the~ID generated from the following general closed form solution for the first few modes~\cite{Rin25}
\begin{align}\label{eq:psi_00_10_22_analytical}
\psi(T,R,\theta,\phi) & = \frac{f(T+R) - f(T-R)}{R} + \bigg[ \frac{1}{R^2} [f(T+R) - f(T-R)] - \frac{1}{R} [f'(T+R) \non \\
& + f'(T-R)] \bigg] \cos\theta +  \bigg[ \frac{3}{R^3} [f(T+R) - f(T-R)] - \frac{3}{R^2} [f'(T+R) \non \\
& + f'(T-R)] + \frac{f''(T+R) - f''(T-R)}{R} \bigg] \sin^2\theta \sin 2\phi \, 
\end{align}
with~$f(x) = e^{-9x^2}$. Fig.~\ref{fig:energy_norm_disc_cont} shows an excellent agreement between the two norms for all time~$t$, both in the~SBP-TEM and~SBP-Stable discretizations, at the resolution of~$(N_r, N_\theta, N_\phi) = (200, 50, 50)$. Meanwhile, Fig.~\ref{fig:energy_norm_disc_cont_diff} illustrates the differences in both linear and logarithmic scales, and also the difference in energies between the two~SBP schemes, thereby quantifying their reliability for long times.

\begin{figure*}[t]
\centering
\includegraphics[width=\textwidth]{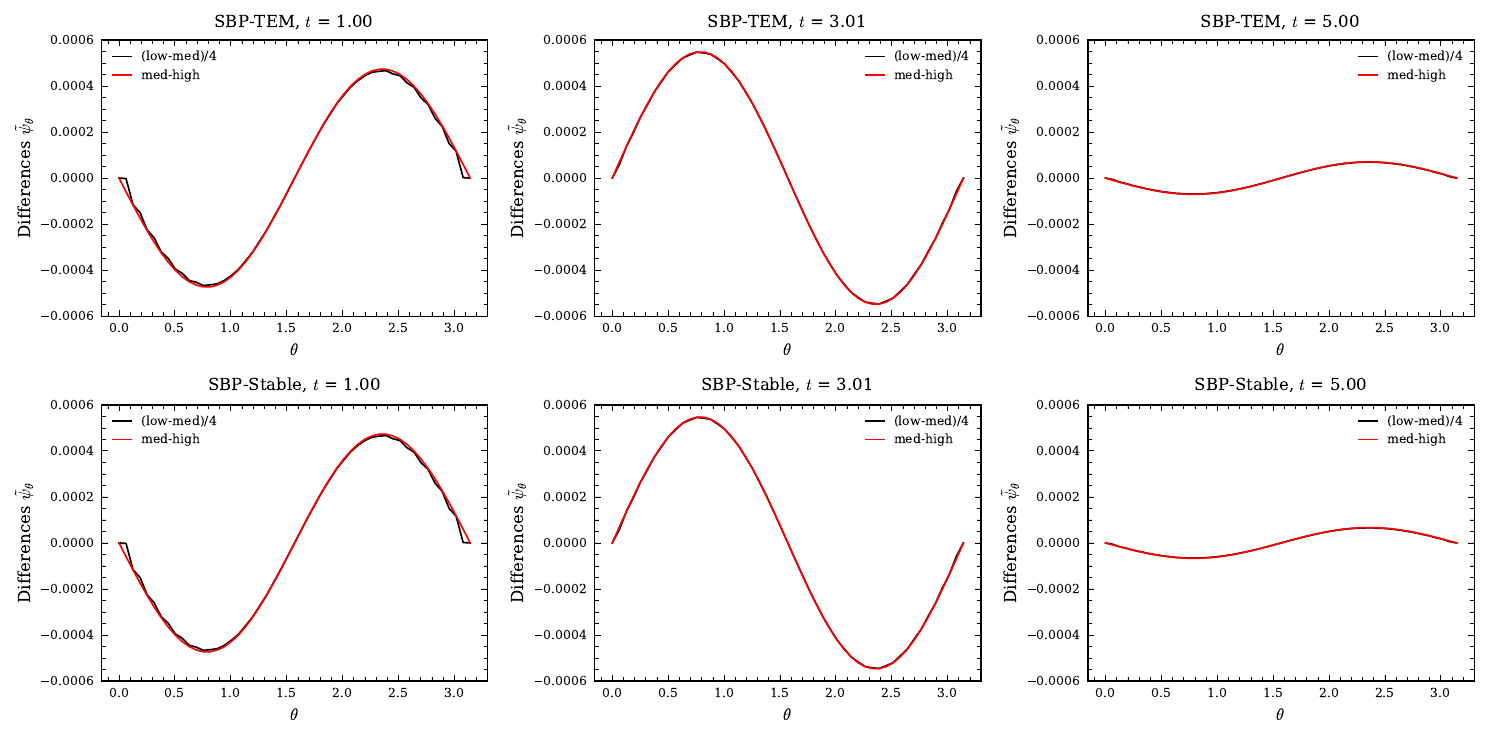}
\caption{Pointwise convergence of~$\tilde{\psi}_{\theta}$ along~$\hat{\theta}$, integrated over~$r$ and~$\phi$, both in~SBP-TEM and~SBP-Stable discretizations, demonstrating a perfect second-order convergence throughout the domain, despite coordinate singularities on the polar axis.}
\label{fig:theta_convergence_psitheta}
\end{figure*}

\begin{figure*}[t]
\centering
\includegraphics[width=\textwidth]{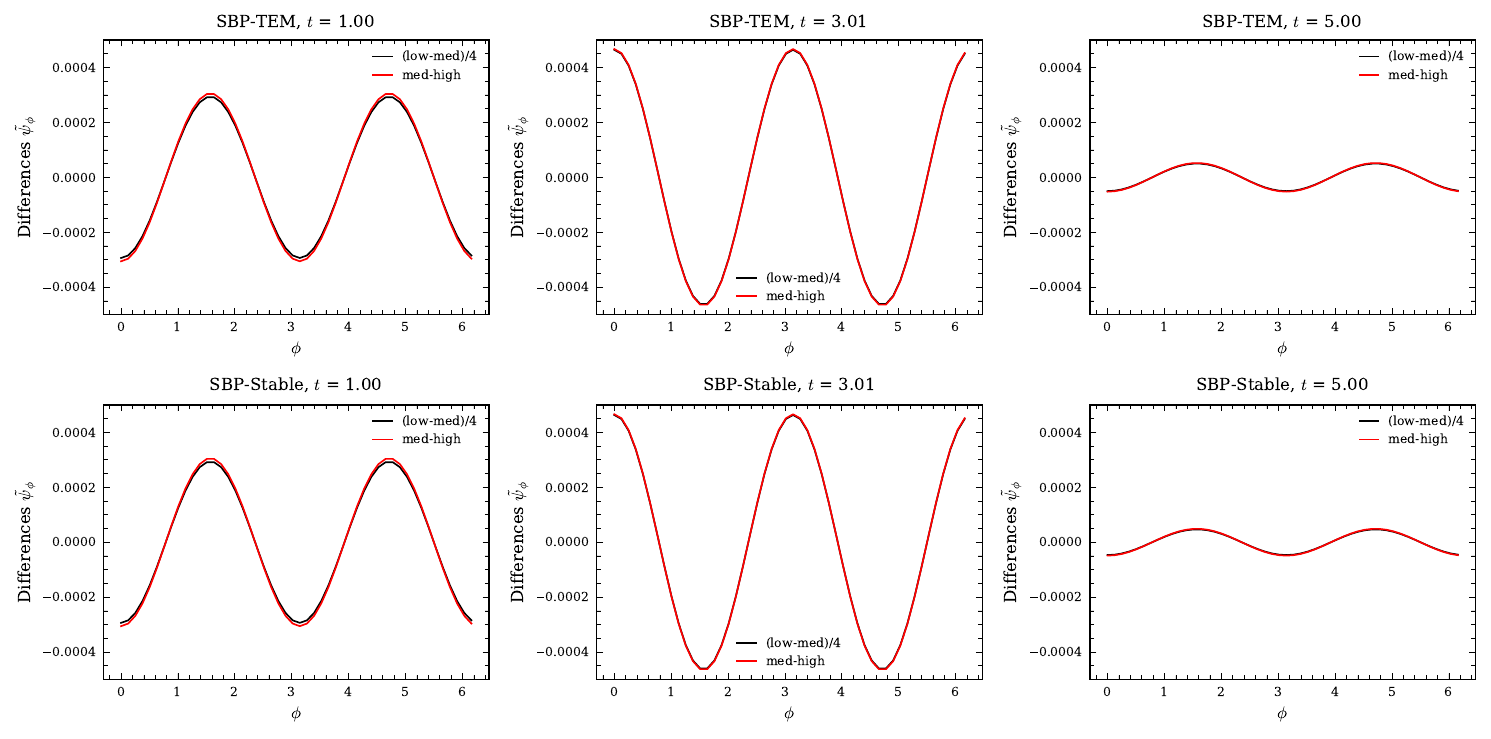}
\caption{Pointwise convergence of~$\tilde{\psi}_{\phi}$ along~$\hat{\phi}$, integrated over~$r$ and~$\theta$, both in~SBP-TEM and SBP-Stable schemes, demonstrating a perfect second-order convergence across the periodic domain~$\phi \in [0, 2\pi)$.}
\label{fig:phi_convergence_psiphi}
\end{figure*}

The left panel in Fig.~\ref{fig:norm_conv_TEM_Stable} shows convergence order in the energy norm for both the SBP-TEM and SBP-Stable schemes over time. These convergence orders oscillate around~$n_{conv} = 2$ due to higher-order contributions to the numerical errors and the very low base resolution~$(N_r, N_\theta, N_\phi) = (50, 8, 16)$. However, these fluctuations diminish as the base resolution increases. These norms are obtained by restricting the variables at all these resolutions to the base resolution.

To encompass all grid points across various resolutions, we define the following norms of errors using the ceiling function~$\lceil \cdot \rceil$,
\begin{align}\label{eq:E1_norm}
E_1(t) & = \sum_{(I,J,K) = (0,0,0)}^{(2 N_r, 2 N_\theta, 2 N_\phi-1)} \frac{\Delta r_I \Delta \theta_J \Delta \phi_K}{8(1 + \delta_{I,N_r})(1 + \delta_{J,N_\theta})} W^\alpha_{I,J,K} \non \\
& \bigg[ \frac{1}{8} \sum_{(I',J',K') = (I-1,J-1,K-1)}^{(I,J,K)} U^L_\alpha \left( \left\lceil \frac{I'}{2} \right\rceil, \left\lceil \frac{J'}{2} \right\rceil, \left\lceil \frac{K'}{2} \right\rceil \right) - U^H_\alpha (I,J,K) \bigg]^2 \, ,
\end{align}
and
\begin{align}\label{eq:E2_norm}
E_2(t) & = \sum_{(I,J,K) = (0,0,0)}^{(2 N_r, 2 N_\theta, 2 N_\phi-1)} \frac{\Delta r_I \Delta \theta_J \Delta \phi_K}{8^2(1 + \delta_{I,N_r})(1 + \delta_{J,N_\theta})} W^\alpha_{I,J,K} \non \\
& \sum_{(I',J',K') = (I-1,J-1,K-1)}^{(I,J,K)} \bigg[ U^L_\alpha \left( \left\lceil \frac{I'}{2} \right\rceil, \left\lceil \frac{J'}{2} \right\rceil, \left\lceil \frac{K'}{2} \right\rceil \right) - U^H_\alpha (I,J,K) \bigg]^2 \, ,
\end{align}
where,~$U_\alpha = \{ \Psi, \Psi_+, \Psi_-, \Psi_\theta, \Psi_\phi \}$,~$W^\alpha$ are the corresponding weights, while~$\delta_{I,J}$ are the Kronecker deltas. In these norms, we observe the same behavior as noted previously. It is important to highlight that~$E_2$ is more sensitive to the higher-order contributions to the numerical errors.

\begin{figure*}[t]
\centering
\includegraphics[width=1.0\linewidth]{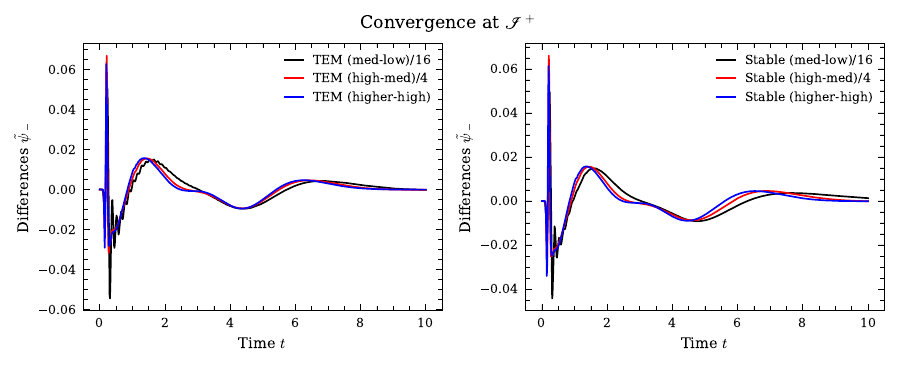}
\caption{Convergence at~$\mathscr{I}^+$ of the outgoing mode~$\tilde{\psi}_-$ integrated over the~$2-$sphere, both in the~SBP-TEM and~SBP-Stable discretizations, demonstrating a second-order convergence with higher-order wiggles that diminish with increasing resolution.}
\label{fig:scri_convergence_psiminus}
\end{figure*}

\begin{figure*}[t]
\centering
\includegraphics[width=1.0\linewidth]{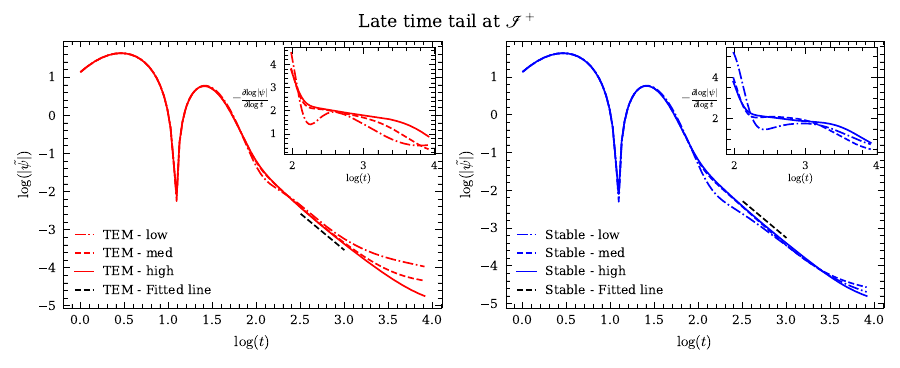}
\caption{Late-time tail at~$\mathscr{I}^+$ for the case~$F = 1/\chi^2$ at three progressively refined resolutions of the rescaled field~$\tilde{\psi}$ in log--log scale, using the natural log~$\ln$. The late-time signal clearly exhibits a power-law decay with slopes shown in the inset plots, which converge to a constant limit as the resolution increases.}
\label{fig:late_time_tail_psi}
\end{figure*}
 
The right panel of Fig.~\ref{fig:norm_conv_TEM_Stable} shows norm convergences achieved by increasing the resolution in just one direction at a time. The base resolutions used for the~$r$-,~$\theta$- and~$\phi$- convergences are~$(N_r, N_\theta, N_\phi) = (200, 20, 20)$,~$(100, 50, 20)$ and~$(100, 20, 50)$, respectively. Pointwise convergence along the~$\hat{r}$,~$\hat{\theta}$ and~$\hat{\phi}$ directions is depicted in Figs.~\ref{fig:r_convergence_psim},~\ref{fig:theta_convergence_psitheta} and~\ref{fig:phi_convergence_psiphi}, respectively, each with its corresponding base resolution specified. In each case, the convergence is integrated along the other two directions. Additionally, convergence at~$\mathscr{I}^+$, integrated over the angles, is plotted over time Fig.~\ref{fig:scri_convergence_psiminus}, with a base resolution of~$(50, 8, 16)$.

\subsection{Scattering Potential, $F=1/\chi^2$}\label{sec:F=1/chi^2}

The presence of a scattering potential leads to the formation of late-time tails at~$\mathscr{I}^+$, following Price's law~\cite{Pri72}. This is illustrated in Fig.~\ref{fig:late_time_tail_psi}, where we show results at different resolutions, starting with~$(N_r, N_\theta, N_\phi) = (50, 8, 16)$. The nested plots demonstrate how the slopes change over time. We anticipate that these slopes will asymptote to~$-2$ if we achieve sufficient resolution and allow enough time~$t$ to pass.

\begin{figure*}[t]
\centering
\includegraphics[width=1.0\textwidth,
trim = {4.2cm 2.8cm 3cm 3.325cm}, clip]
{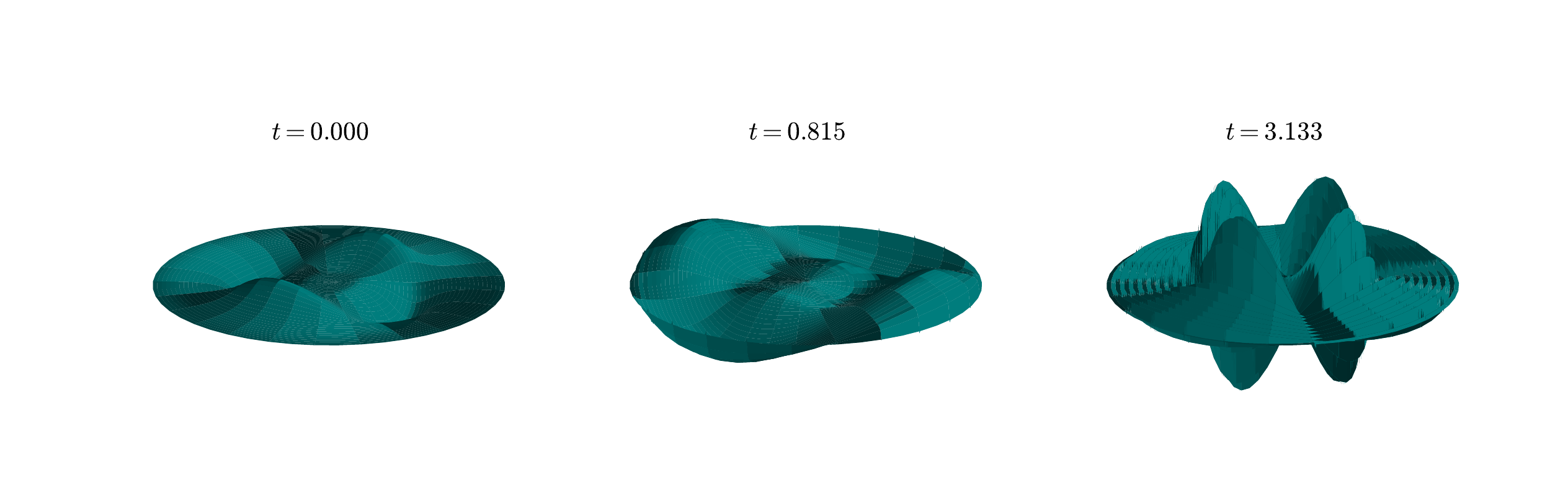}
\caption{Propagation of the numerical solution to the massive Klein-Gordon equation on hyperboloidal slices, reflecting back and forth along the radial direction ensure total energy is conserved at all times.}
\label{fig:m2_evolution}
\end{figure*}

\begin{figure}[t]
\centering
\includegraphics[width=0.6\linewidth]{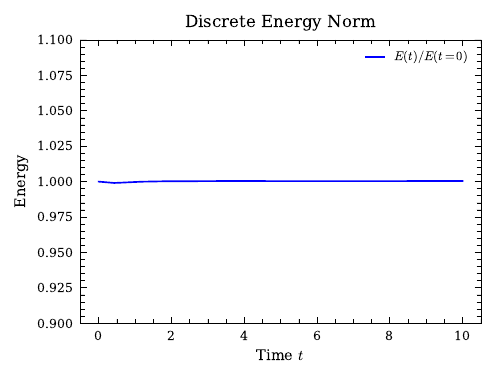}
\caption{Energy conservation of the numerical solution over time for the case~$F=m^2$ on hyperboloidal slices in the SBP–Stable scheme with zero dissipation.}
\label{fig:m2_energy_norm_zero_dissipation}
\end{figure}

\begin{figure*}[t]
\centering
\includegraphics[width=1.0\linewidth]{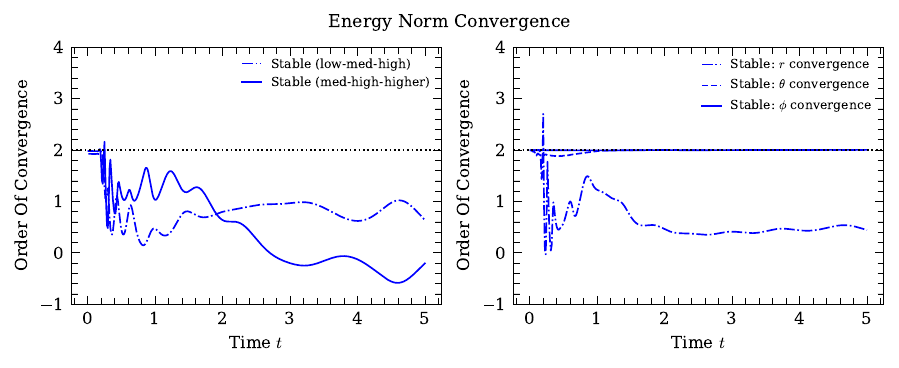}
\caption{Convergence order in the energy norm for the massive case~$F=m^2$. \emph{Left panel}: Convergence with uniform refinement along all directions. \emph{Right panel}: Convergence obtained by refining one coordinate direction at a time. While the angular directions demonstrate perfect second-order convergence, the convergence in the radial direction is compromised due to the singular behavior at~$\mathscr{I}^+$.}
\label{fig:m2_energy_norm_convergence_Stable}
\end{figure*}

\begin{figure*}[t]
\centering
\includegraphics[width=1.0\linewidth]{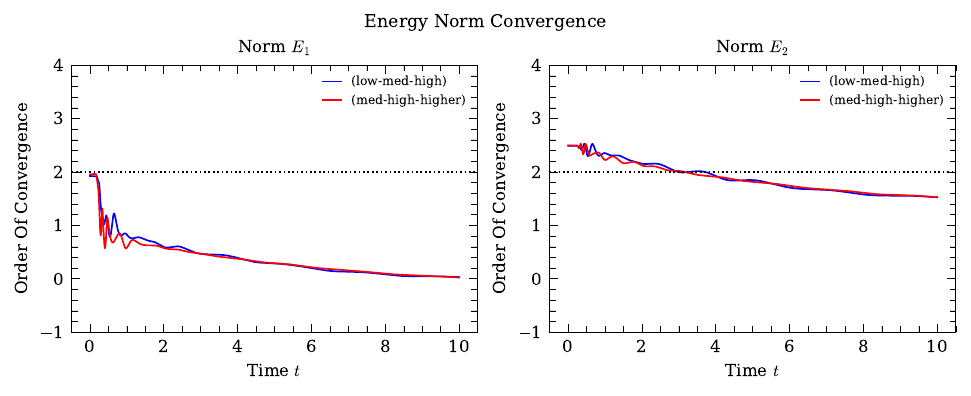}
\caption{Convergence order in the energy norm for~$F=m^2$, with norms given in~\eqref{eq:E1_norm} and~\eqref{eq:E2_norm}. The plots demonstrate a perfect correlation in the convergence orders across different base resolutions.}
\label{fig:m2_energy_norm_convergence_Stable_E1_E2}
\end{figure*}

\subsection{Massive Klein-Gordon Equation, $F=m^2$}\label{sec:F=m^2}

The case~$F = m^2$ makes the potential terms singular at~$\mathscr{I}^+$, both in the equations and the energy norm. However, because the solutions decay sufficiently rapidly, as demonstrated in previous studies~\cite{Kla93, Win88, GauVanHil21}, the signal never actually reaches~$\mathscr{I}^+$, and remains confined within the light cone. Despite this singularity, the system evolves in a stable manner, and the total discrete energy of the system is conserved at all times when dissipation is turned off, as illustrated in Figs.~\ref{fig:m2_evolution} and~\ref{fig:m2_energy_norm_zero_dissipation}.

Switching on artificial dissipation removes the high-frequency modes; however, the convergence order decreases over time, as illustrated in the left panel of Fig.~\ref{fig:m2_energy_norm_convergence_Stable}. Since the singularity exists only in the radial direction, the convergence properties in the angular directions remain unchanged, as shown in the right panel of Fig.~\ref{fig:m2_energy_norm_convergence_Stable}. This observation is consistent with the Lax-Equivalence theorem~\cite{Tho95}.

The left plot in Fig.~\ref{fig:m2_energy_norm_convergence_Stable} indicates a loss of correlation between the convergence orders at lower and higher resolutions after a certain period. This occurs because the convergence orders are computed on a coarser grid. However, we do not observe this loss of correlation in the convergence orders derived from the $E_1$ and $E_2$ norms, as defined in \eqref{eq:E1_norm} and \eqref{eq:E2_norm}, respectively. This is illustrated in Fig.~\ref{fig:m2_energy_norm_convergence_Stable_E1_E2}. Since these norms integrate information from all grid points across all resolutions, Fig.~\ref{fig:m2_energy_norm_convergence_Stable_E1_E2} highlights the advantages of these norms compared to conventional ones.

\section{Conclusions}\label{sec:conclusions}

This work presents the complete three-dimensional (3D) SBP framework on hyperboloidal slices within a Minkowski background, as developed in~\cite{GauVanHil21, GauRedKum26} with~$\mathscr{I}^+$-fixing coordinates. We begin by redefining the linear wave equation (LWE) at the coordinate singularities, specifically at the $z$-axis and the origin. Following this, we introduce the first-order reduction (FOR) in a covariant manner. This is succeeded by the introduction of compactified hyperboloidal coordinates, along with a rescaling factor associated with this compactification, given by~\eqref{eq:Chi_natural}. Interestingly, this rescaling not only reduces to the conformal rescaling in the case of conformal compactification, but it also significantly simplifies the regularized covariant divergence operators, as defined in~\eqref{eq:Tilded_Operators} and~\eqref{eq:Tilded_Operator_simplified}.

The FOR system is further modified by adding constraint damping terms to the spatial part via vector addition, and introducing artificial dissipation to its temporal part, resulting in a geometric form for the overall modification.

The overall SBP scheme is derived by discretizing the partial derivatives and the regularized covariant divergence operators. We impose energy conservation, as discussed in equations \eqref{eq:E-dot_rescaled} and \eqref{eq:Stokes_diff_form_rescaled}. In Section \ref{sec:Dissipation}, we also introduce suitable dissipation operators in spherical-polar coordinates. These operators are defined everywhere, including at the boundary points, and they satisfy the dissipative property (DP) in terms of energy norms.

We derive two types of SBP schemes: SBP-TEM and SBP-Stable, that differ only in their treatment of the outer boundary. Each has its own strengths and weaknesses. The SBP-TEM scheme maintains the accuracy of the numerical solution throughout the entire domain, while the SBP-Stable scheme guarantees stability and negative-definiteness of the energy flux at~$\mathscr{I}^+$. We illustrate these properties through a series of numerical tests. Although the SBP-TEM scheme demonstrates superior convergence properties, the SBP-Stable discretization is capable of effectively evolving systems with singularities, such as the massive Klein-Gordon fields discussed in Sec.~\ref{sec:F=m^2}. Both schemes excel at managing coordinate singularities in curvilinear coordinates while sustaining convergence properties similar to those of completely regular systems. This capability is one of the notable strengths of this approach.

We focused on second-order accurate finite-difference (FD) methods in this study. However, the scheme outlined in Section \ref{sec:SBP_Scheme} can also be adapted for higher-order accurate FD methods and pseudo-spectral discretizations. Even at low resolutions and with limited accuracy, this scheme effectively captured the essential physical properties of the corresponding continuum systems. For example, it successfully represented wave propagation to~$\mathscr{I}^+$ for the case~$F = 0$, cf. Fig.~\ref{fig:Contour_TEM}, late-time tails at~$\mathscr{I}^+$ for systems with scattering potentials that decay sufficiently fast asymptotically, cf. Fig.~\ref{fig:late_time_tail_psi}, and demonstrated energy conservation in the massive Klein-Gordon case, cf. Fig.~\ref{fig:m2_evolution}.

We also propose new norm convergence tests, defined by~\eqref{eq:E1_norm} and~\eqref{eq:E2_norm}, which utilize data from all grid points across all resolutions. These tests not only highlight higher-order contributions to numerical errors more effectively but also provide improved convergence results. This is particularly evident in scenarios where traditional norm convergence tests, which rely on data from coarser grids, tend to fail. The effectiveness of our new tests is illustrated in Figs.~\ref{fig:m2_energy_norm_convergence_Stable} and \ref{fig:m2_energy_norm_convergence_Stable_E1_E2}.

\section{Acknowledgements}\label{sec:Acknowledgements}

This work was partially supported at the ``Extremal Black Holes and the Third Law of Black Hole Thermodynamics'' workshop held at the Institute for Computational and Experimental Research in Mathematics (ICERM) at Brown University. SG thanks Jeffrey Winicour, Thomas Maedler, Radouane Gannouji and Steven Liebling for the invitation to write this summary. All the simulations were executed on the Sonic cluster at the International Centre for Theoretical Sciences (ICTS) of the Tata Institute of Fundamental Research (TIFR), Bengaluru. SG's research was partially supported by the Beijing National Science Foundation~(BJNSF) under the International Scientist Project~(ISP), Grant No.: IS25026.

\bibliography{Refs}

\appendix

\end{document}